\documentclass[10pt]{article}
\usepackage{fancyhdr}
\usepackage{extramarks}
\usepackage{amsmath}
\usepackage{amsthm}
\usepackage{amsfonts}
\usepackage{siunitx}
\usepackage{tikz}
\usepackage[plain]{algorithm}
\usepackage{algpseudocode}
\usepackage{multirow}
\usepackage{booktabs}
\usepackage{graphicx}
\usepackage{subfigure}
\usepackage[colorlinks,linkcolor=black,anchorcolor=black,citecolor=black,urlcolor=blue]{hyperref}
\usepackage{amsmath,bm}
\usepackage{booktabs}
\usepackage{mathtools}
\usepackage{amssymb}
\usepackage{caption}
\usepackage{capt-of}
\usepackage{mciteplus}
\usepackage{cite}
\usepackage{mathrsfs}
\usepackage[title,titletoc,toc]{appendix}
\usepackage{xr}
\usepackage{parskip}
\usepackage{soul}
\usepackage{textcomp}
\usepackage[colaction]{multicol}
\usepackage[switch]{lineno}
\usepackage{lipsum}
\usepackage{etoolbox}
\usepackage{longtable}
\usepackage{array}
\usepackage{tablefootnote}
\usepackage{ragged2e}
\usepackage{soul}
\newcolumntype{C}[1]{>{\centering\arraybackslash}p{#1}}
\usetikzlibrary{automata,positioning}
\usetikzlibrary{automata,positioning}

\usepackage{xr}

\begin{document}
	\title{K-Nearest-Neighbors Induced Topological PCA for Single Cell RNA-Sequence Data Analysis}
	
	\author{Sean Cottrell$^1$, Yuta Hozumi$^1$ and  Guo-Wei Wei$^{1,2,3}$\footnote{
			Corresponding author.		Email: weig@msu.edu} \\
		\\
		$^1$ Department of Mathematics, \\
		Michigan State University, East Lansing, MI 48824, USA.\\
		$^2$ Department of Electrical and Computer Engineering,\\
		Michigan State University, East Lansing, MI 48824, USA. \\
		$^3$ Department of Biochemistry and Molecular Biology,\\
		Michigan State University, East Lansing, MI 48824, USA. \\
	}
	\date{\today} 
	
	\maketitle
	
	\begin{abstract}
	
	Single-cell RNA sequencing (scRNA-seq) is widely used to reveal heterogeneity in cells, which has given us insights into cell-cell communication, cell differentiation, and differential gene expression. However, analyzing scRNA-seq data is a challenge due to sparsity and the large number of genes involved. Therefore, dimensionality reduction and feature selection are important for removing spurious signals and enhancing downstream analysis. 
	Traditional PCA, a main workhorse in dimensionality reduction,  lacks the ability to capture geometrical structure information embedded in the data, and previous  graph Laplacian regularizations are limited by the analysis of only a single scale.  
	We propose a  topological  Principal Components Analysis (tPCA) method by the  combination of persistent Laplacian (PL) technique and L$_{2,1}$ norm regularization to address multiscale and multiclass heterogeneity issues in data.  
 We further introduce a k-Nearest-Neighbor (kNN) persistent Laplacian technique to improve the robustness of our persistent Laplacian method.  
The proposed kNN-PL is a new algebraic topology technique which addresses the many limitations of the traditional persistent homology. 
  Rather than inducing filtration via the varying of a distance threshold, we introduced kNN-tPCA, where filtrations are achieved by varying the number of neighbors in a kNN network at each step, and find that this framework has significant implications for hyper-parameter tuning. We validate the efficacy of our proposed tPCA and kNN-tPCA methods on 11 diverse benchmark scRNA-seq datasets, and showcase that our methods outperform  other unsupervised PCA enhancements from the literature, as well as   popular Uniform Manifold Approximation (UMAP), t-Distributed Stochastic Neighbor Embedding (tSNE), and Projection Non-Negative Matrix Factorization (NMF) by  significant margins. For example, tPCA provides up to  628\%, 78\%, and  149\% improvements to UMAP, tSNE, and NMF, respectively on classification in the F1 metric, and kNN-tPCA offers 53\%, 63\%, and  32\% improvements to UMAP, tSNE, and NMF, respectively on clustering in the ARI metric. 
	
	\end{abstract}
keywords: {Topology, Persistent Homology, Persistent Laplacian, scRNA-seq, dimensionality reduction, clustering, machine learning}
	
	\newpage
	
	{\setcounter{tocdepth}{4} \tableofcontents}
	\setcounter{page}{1}
 
	\newpage
	
	\section{Introduction}
	Single cell RNA sequencing (scRNA-seq) is a relatively new method that profiles transcriptomes of individual cells, revealing vast information in the heterogeneity within cell population, which has lead to further understanding of gene expression, gene regulation, cell-cell communication, cell differentiation, spatial transcrtiptomics, signal transduction pathways, and more \cite{lun2016step, kharchenko2021triumphs}. The workflow of a typical scRNA-seq analysis involves single cell isolation, RNA extraction and sequencing using a library and downstream analysis. With the technological improvements, more than 20,000 genes can be profiled, which has led to a high-dimensionality challenge. Despite the improvements in the methodology that allows for more accurate reading of genes and increasing the number of sequenced cells per experiment, analyzing the data for downstream analysis remains a hurdle. Numerous methods and procedures have been proposed to analyze the data \cite{luecken2019current, chen2019single, petegrosso2020machine, li2019statistical, andrews2021tutorial, lahnemann2020eleven, flores2022deep}. Specific challenges in scRNA-seq data analysis include  drop-out events-induced zero expression read count, inadequate sequencing depth-induced  inconsistent low reading counts, noise data, and high dimensionality  \cite{lahnemann2020eleven, jiang2022statistics}. Therefore, dimensionality reduction and feature selection to eliminate low signals is an essential step in analyzing scRNA-seq data.

    Various dimensionality reduction and feature selection methods have been proposed for analyzing  scRNA-seq data. ScRNA by non-negative and low rank representation (SinLRR) assumes that scRNA-seq data is inherently low rank and finds the smallest ranked matrix that approximates the original data \cite{zheng2019sinnlrr}. Single-cell interpretation via multikernel learning (SIMLR) utilizes multiscale kernel to learn a cell-cell similarity metric that can be used for downstream analysis \cite{wang2017visualization}. Deep learning has also been used to perform dimensionality reduction \cite{flores2021deep, 10.3389/fgene.2021.733906, article, article2}.

    Traditional dimensionality has also been widely incorporated into scRNA-seq analysis pipeline. Non-linear dimensionality reduction, such as uniform manifold approximation and projection (UMAP), t-distributed stochastic neighbor embedding (t-SNE), multidimensional scaling (MDS) and isomap have been utilized for visualization \cite{JMLR:v9:vandermaaten08a, mcinnes2020umap, article3, inproceedings}. However, directly applying such algorithm can be challenging because these methods rely on distance calculation  and  data sparsity, but high dimensional scRNA-seq data may suffer from poor distance calculations. Recently, correlated clustering and projection (CCP) has been used  on scRNA-seq data and its visualization \cite{hozumi2022ccp,hozumi2023preprocessing}. CCP utilizes gene-gene correlation to partition genes, and uses cell-cell correlation on the partitioned genes to project the original genes into super-genes. Non-negative matrix factorization (NMF) has been widely utilized due to its interpretability. NMF uses matrix factorization, where the basis can be interpreted as meta-gene, which are weighted sums of the original genes. Numerous NMF with various constrains has been proposed for scRNA-seq \cite{doi:10.1021/acs.jcim.2c01305, article4}. One of the oldest dimensionality reduction methods, principal components analysis (PCA) is still one of the most popular method used for scRNA-seq \cite{jolliffe2005principal}.

    PCA is a linear dimensionality reduction algorithm, where the goal is to compute a orthonormal basis that maximizes the variance of the projected data. The first component is call the principle component, where the variance of the projected data is maximized. The subsequent components $i$ are orthogonal to the $i-1$ components and maximizes the variance of the residual of the approximation. Though PCA is popular because of its efficiency and is linear due to the having no reliance on utilizing metric computation,  it has its drawbacks. First, PCA lacks concrete interpretability because of its eigenmode-representation the data. In scRNA-seq, there are many 0-counts in the data, i.e. the gene is not expressed, which can contribute to the principle component and the coefficient matrix can be   dense. Second, PCA assumes that the noise of the data is Gaussian, which may not model scRNA-seq data well because the original data is a count matrix. To tackle the first problem, sparse PCA (sPCA) was introduced  \cite{inproceedings}, where by adding a $l_1$ penalty term on the basis, it allowed for sparsity in the principle components. An alternative formulation of PCA was introduced by Nie et. al. \cite{nie2016nongreedy} to improve robustness of PCA by assuming the noise is sampled from the Laplace distribution, which is called rPCA. Graph regularization was further introduced by Jiang et. al. which utilizes a graph Laplacian to incorporate nonlinear manifold structure to the reduction \cite{jiang2013graph}. However, graph Laplacian can only capture a single scale, and is not able to capture the topological structure of the data.


    We   introduced persistent Laplacian-Enhanced PCA theory in previous works, which was shown to outperform all other state-of-the-art PCA enhancements for microarray data analysis \cite{cottrell2023plpca}.  Persistent Laplacian, also called persistent combinatorial Laplacian or persistent spectral graph, was introduced as a new generation of topological data analysis (TDA) methods in 2019 \cite{wang2020persistent}. It has stimulated a variety of theoretical developments \cite{memoli2022persistent,wei2021persistent,liu2023algebraic,chen2023persistent} and led to remarkable applications \cite{chen2022persistent,qiu2023persistent,meng2021persistent}.     
		 PLPCA has the ability to recognize the stability of topological features in our data at multiple scales, and provide a more thorough spatial view through the filtration of a simplicial complex, which induces a sequence of simplicial complexes. Accounting for the spectra of each corresponding Laplacian matrix for each complex in the sequence enables us extract this topological information, improving our ability to preserve intrinsic geometrical information during dimensionality reduction. We can accomplish this by constructing a weighted sum of each Laplacian matrix, generating an accumulated spectral graph. However, PLPCA requires intensive parametrization. 
		
		The objective of the present work is 	to introduce L$_{2,1}$ norm regularization to improve the sparsity and  heterogeneity constraint in PLPCA. The resulting technique, called  topological PCA (tPCA), allows us to significantly reduce the distribution of the weightings in the accumulated spectral graph, while still achieving near optimal performance.
Additionally, we introduce a k-nearest neighbor (kNN) persistent Laplacian algorithm to improve the robustness of our tPCA. The performance of resulting kNN-tPCA does not depend on parameter search.   We extensively validate the performance of tPCA and kNN-tPCA  for clustering and classification on a series of 11 scRNA-seq datasets. We demonstrate that our new method is superior to other PCA enhancements as well as NMF. 

    Our work then proceeds as follows. First, we review the mathematical formulations of each of the enhancements that build up to tPCA. Specifically, graph regularization and sparseness. We then discuss the tools for perssistent Laplacian theory which we will incorporate to arrive at our final procedure: RpLSPCA, or tPCA. We then validate the performance of this new method on a set of 11 scRNA-seq datasets against other notable PCA enhancements, as well as NMF, for clustering and classification. Extensive tests indicate that our methods are the state of the art procedure for dimensionality reduction prior to clustering and classification. Specifically, over the 11 tested datasets, tPCA outperforms UMAP on average by 628\%  on the F1 metric. Lastly, we visualize the tPCA Eigen-Genes via UMAP and tSNE, as well as Residue Similarity Plots to further assess the performance of our methods.

\section{Methods}
In this section, we provide an overview of PCA and its derivatives, including sparse PCA (sPCA), and graph Laplacian regularized sparse PCA (gLSPCA). We then introduce our method, topological PCA (tPCA) and a less parameter-intensive alternative kNN-tPCA.
 We will first define the notation used in the following subsection.

\subsection{Notations} 
We summarize our notations as follows.  
\begin{enumerate}
    \item $X = \{\mathbf{x}_1, ..., \mathbf{x}_N\} \in \mathbb{R}^{M\times N}$, where $\mathbf{x}_j \in \mathbb{R}^M$ indicates the $j$th sample or cell, and $M$ is the number of genes.
    \item $\|A\|_F = \sqrt{\sum_{j=1}^N\sum_{i=1}^M A_{ij}}$ is the frobenius norm of matrix $A$.
    \item $\|A\|_{2,1} = \sum_{j=1}^N \|\mathbf{a}_j\|_2$ is the $l_{2,1}$ norm of $A$, where the   summation is taken after taking the $l_2$ norm of each column. Alternatively, we can think of this as taking the $l_2$-norm of the genes and taking the sum over the cells.
    \item $\text{Tr}(A)$ is the trace of matrix $A$.
    \item Let $m << M$ the dimension of the subspace, and $M$ is the number of original dimension, ie the number of genes.
    \item Let $N$ be the number of samples or cells.
    \item $U \in \mathbb{R}^{m \times M}$ is the principle components, or the basis of the lower dimensional subspace of $X$, and $m$ is the number of dimension.
    \item $Q \in \mathbb{R}^{m \times N}$ is the projection of $X$ onto the subspace spanned by $U$.
\end{enumerate}
 
\subsection{PCA}
Principal Component Analysis has historically served as the baseline approach for dimensionality reduction during gene expression data analysis \cite{jolliffe2016principal}. The goal of PCA is to express some high dimensional data $X \in \mathbb{R}^{M \times N}$ in a lower dimensional space. This is accomplished via computing the principal components, which are the eigenvectors of the covariance of $X$ corresponding to the largest eigenvalues. Alternatively, we can express PCA as finding a $m$-dimensional subspace that approximate the data matrix, i.e.,
	\begin{align}
		\min_{U, Q}\|X - UQ^T\|_{F}^2, \quad \text{s.t. } Q^TQ = I_N
	\end{align}
where $Q^TQ = I_N$ is the orthonormal constrain, and $I_N$ is the $ N\times N$, or cell by cell, identity matrix. When the original matrix $X$ is 0-mean 1-variance scaled, we can see that this formulation is equivalent to take the eigenvectors of the covariance matrix. Alternatively, the orthonormal constrain can be applied to $U$, which yields traditional PCA.

%
 
\subsection{Sparse PCA}
PCA requires that the principal components be expressed as linear combinations of all the features with non-zero weightings. However, in the context of gene expression data analysis, this introduces unnecessary computational complexity and noise, because many genes are not expressed in the cells, or is only expressed under particular circumstances \cite{luecken2019current}. Therefore, the interpretability of PCA is significantly aided by the introduction of Sparse PCA (sPCA), which allows for zero weightings \cite{inproceedings}. Sparse PCA can take several forms, notably the inclusion of an $L_{2,1}$ norm penalty term in the objective function as in Eq. \ref{eq:spca}:
\begin{align}\label{eq:spca}
    \min_{U,Q}\|X - UQ^T\|_F + \beta\|Q\|_{2,1}, \quad \text{s.t. } Q^TQ = I_N
\end{align}
The $L_{2,1}$ norm is defined as $\|A\|_{2,1} = \sum^n_{i=1} \| {\mathbf{a}}_i\|_2$, or first calculating the L$_2$ norm of each row, and then computing the L$_1$ norm of row-based  L$_2$ norms.  The $\beta$ parameter scales the sparse regularization term. 
	
\subsection{Graph Laplacian Regularized Sparse PCA}
While the inclusion of sparse regularization should address some of the shortcomings of PCA relating to interpretability, it does not address the inability of PCA in recognizing complex geometric structures which are present in the higher dimensional space. Graph Laplacian has been commonly used to incorporate geometric information into the reduction such that similar samples in high dimension will be closer in the lower dimensional embedding.  This can be accomplished with neighbor graphs with pairwise edges, specifically, the Laplacian operator \cite{belkin2001laplacian}.

Let $G(V, E, W)$ be a nearest neighbor graph, where $V$ is the set of vertices, $E$ is the edge, and $\omega$ are the weights of the edges. $E$ can be defined as $E = \{(\mathbf{x}_j, \mathbf{x}_i): \mathbf{x}_i \in \mathcal{N}_k(\mathbf{x}_j) \text{ or } \mathbf{x}_j \in \mathcal{N}_k(\mathbf{x}_i)\}$, where $\mathcal{N}_k(\mathbf{x}_j)$ is the $k$-nearest neighbors of sample $j$ under some metric. For pairs of points in the edge set, we can define a weight satisfying the following two properties
\begin{align*}
    & \Phi(\mathbf{x}_i, \mathbf{x}_j) \to 1 \quad  \text{as} \|\mathbf{x}_i - \mathbf{x}_j\| \to 0 \\
    & \Phi(\mathbf{x}_i, \mathbf{x}_j) \to 0 \quad \text{as} \|\mathbf{x}_i - \mathbf{x}_j\| \to \infty.
\end{align*}
Such condition is satisfied by a class of functions called radial basis functions. For this work, we utilize the Gaussian kernel as the edge weights
\begin{equation}\label{eq:W}
W_{ij} = 
    \begin{cases} 
    e^{- \| \mathbf{x}_i, \mathbf{x}_j\|^2/\eta} & \text{ if }\mathbf{x}_j \in \mathcal{N}_k(\mathbf{x}_i) \\
    0, & \text{ otherwise}. \\
    \end{cases}
\end{equation}

The matrix $W$ is known as the weighted adjacency matrix. Here, $\eta \in \mathbb{R}$ defines the geodesic distance, or the width of the Gaussian kernel. We can then define the Graph Laplacian $L$, by taking
\begin{align}
    L = D - W,\\
    D_{ii} = \sum_{j=1}^N W_{i,j},    
\end{align}
where $D$ is the degree matrix, defined as the row sum of $W$, which shows the total connectivity of the vertex $i$. Laplacian graph provides a graphical embedding, which can be used as a regularization for PCA \cite{jiang2013graph}.

In order to incorporate manifold regularization into the PCA framework, consider the distance $\|\mathbf{q}_i - \mathbf{q}_j\|^2$, where $\mathbf{q}_i$ and $\mathbf{q}_j$ correspond to the lower dimensional representation of samples $\mathbf{x}_i$ and $\mathbf{x}_j$, respectively. Using the graph weights $W_{ij}$, we see that if $W_{ij} \to 1$, ie $\mathbf{x}_i$ and $\mathbf{x}_j$ are similar, then $\|\mathbf{q}_i - \mathbf{q}_j\| \to 0$. Alternatively, if $W_{ij} \to 0$, ie $\mathbf{x}_i$ and $\mathbf{x}_j$ are dissimilar, $\|\mathbf{q}_i - \mathbf{q}_j\|^2 \to \infty$. Using this fact, we want to minimize the following.
\begin{align*}
    R & = \frac{1}{2} \sum_{ij}^N W_{ij}\|\mathbf{q}_i - \mathbf{q}_j\|^2  \\ 
      & = \frac{1}{2}\sum_{ij}^N W_{ij} \mathbf{q}_i^T\mathbf{q}_i + \mathbf{q}_j^T\mathbf{q}_j - \sum_{ij}^N W_{ij}\mathbf{q}_i^T\mathbf{q}_j^T \\
      & =  \sum_i^N D_{ii}\mathbf{q}_i^T\mathbf{q}_i - \sum_{ij}^N W_{ij}\mathbf{q}_i^T\mathbf{q}_j^T \\
      & = \text{Tr}(Q^TDQ) - \text{Tr}(Q^TWQ) \\
      & = \text{Tr}(Q^TLQ).
\end{align*}

Utilizing the sparse PCA and the manifold regularization, we obtain the graph Laplacian sparse PCA (gLSPCA)
\begin{align}
    \min_{U,Q}\|X - UQ^T\|_{F} + \beta\|Q\|_{2,1} + \gamma \text{TR}(Q^TLQ), \quad \text{s.t. } Q^TQ = I_N.
\end{align}
 We also observe that the loss function of gLSPCA is Frobenius norm regularization, which is sensitive to outliers and data  heterogeneity when dealing with multiclass data. We then propose replacing Frobenius norm regularization with $\text{L}_{2,1}$ norm regularization to achieve robustness. This yields the following new  objective function, which we call Robust graph Laplacian Sparse PCA (RgLSPCA), 
\begin{align}\label{eq:RgLSPCA}
    \min_{U,Q}\|X - UQ^T\|_{2,1} + \beta\|Q\|_{2,1} + \gamma \text{Tr}(Q^TLQ), \quad \text{s.t } Q^TQ = I_n.
\end{align}

\subsection{Topological PCA}
 
While the inclusion of sparseness and graph Laplacian regularization seeks to address interpretability and geometric structure capture, it still lacks the ability to recognize the stability of topological features at multiple scales, as well as homotopic shape information \cite{wang2020persistent}. To this end, we turn to persistent Laplacian regularization. Like persistent homology, persistent spectral graph theory tracks the birth and death of topological features of   data  as they change over scales \cite{memoli2022persistent, chen2021evolutionary}. We perform this analysis via a filtration procedure on our data to construct a family of geometric structures \cite{wang2020persistent}. We then can study the topological properties of each configuration by its corresponding Laplacian matrix. 

First, we must briefly review the notion of a simplex, simplicial complex, $q$-chain, and boundary. A 0-simplex is a vertex, a 1-simplex is an edge, a 2-simplex is a triangle, and so on. Generally, we consider a $q$-simplex, $\sigma_q$. A simplicial complex is then a means of approximating a topological space by gluing together the faces of simplices. More formally, a simplicial complex $K$ is a collection of simplices such that:
\begin{enumerate}
    \item If $\sigma_q \in K$ and $\sigma_p$ is a face of $\sigma_q$ then $\sigma_p \in K$.
    \item The nonempty intersection of any two simplices is a face of both simplices.
\end{enumerate}
A $q$-chain is then defined as a formal sum of $q$-simplices in a simplicial complex $K$ with coefficients in $\mathbb{Z}_2$. The set of $q$-chains has a basis in the set of $q$-simplices in $K$, and this set forms a finitely generated free Abelian group $C_q(K)$. We then define the boundary operator as a homomorphism relating the Chain groups, $\partial_q : C_q(K) \rightarrow C_{q-1}(K)$. The boundary operator is defined as: 
\begin{equation}
    \partial_q \sigma_q = \sum_{i=0}^q(-1)^i\sigma_{q-1}.
\end{equation}
where $\sigma_{q-1}$ is a $q-1$ simplex. The sequence of chain groups connected by this homomorphism is then a Chain Complex: 
\begin{center}
    $... \xrightarrow{\partial_{q+1}} C_{q}(K) \xrightarrow{\partial_{q}} C_{q-1}(K) \xrightarrow{\partial_{q-1}} ...$.
\end{center}
It is well known that the boundary operator and the Chain Complex associated with a simplicial complex gives the number of $q$-dimensional holes in that topological space. Specifically, the $q$th Homology Group is defined as $H_q = \text{ker} \partial_q / \text{Im} \partial_q$. This is also known as the $q$th Betti Number, $\beta_q$. The matrix representation of the $q$th boundary operator with respect to the standard basis in $C_q(K)$ and $C_{q-1}(K)$ is given as $\mathcal{B}_q$. Besides considering the Homology of our topological space, we can also consider its cohomology. To that end, we define the adjoint operator of $\partial_q$ as: 
\begin{equation}
    \partial_q^*: C_{q-1}(K) \rightarrow C_q(K),
\end{equation}
and the transpose of $\mathcal{B}_q$, denoted $\mathcal{B}_q^T$, is the matrix representation of $\partial_q^*$ with respect to the same basis. We can now define the $q$-combinatorial Laplacian matrix as:
\begin{equation}
    \mathcal{L}_q := \mathcal{B}_{q+1}\mathcal{B}_{q+1}^T + \mathcal{B}_q^T\mathcal{B}_q.
\end{equation}
The harmonic spectrum of the $q$-combinatorial Laplacian matrix reveals the dimension of the $q$th Homology group, or the number of $q$-dimensional holes in our simplicial complex. The non-harmonic spectrum then reveals further homotopic shape information \cite{wang2021hermes}. Intuitively, $\beta_0$ reveals the number of connected components in $K$, $\beta_1$ reveals the number of loops in $K$, and $\beta_2$ reveals the number of 2D voids in $K$. 

However, this framework is confined to the analysis of only a single simplicial complex, or the connectivities at only a single scale. To enrich our spectral information, Persistent spectral graph theory proposes creating a sequence of simplicial complexes by varying a filtration parameter \cite{wang2021hermes}: 
\begin{center}
    $\{ \emptyset \} = K_0 \subseteq K_1 \subseteq ... \subseteq K_p = K$.
\end{center}
For each subcomplex $K_t$ we can denote its chain group to be $C_q(K_t)$, and the $q$-boundary operator $\partial_q^t: C_q(K_t) \rightarrow C_{q-1}(K_t)$. By convention, we define $C_q(K_t) = \{0\}$ for $q<0$ and the $q$-boundary operator to then be the zero map. The boundary operator and adjoint boundary operator are otherwise defined similarly as before for each $K_t$ in the sequence, which allows us to define a sequence of Chain Complexes. 

Next, we introduce persistence to the Laplacian spectra. Define the subset of $C_q^{t+p}$ whose boundary is in $C_{q-1}^t$ as $\mathbb{C}_q^{t,p}$,  assuming the natural inclusion map from $C_{q-1}^t \rightarrow C_{q-1}^{t+p}$.
\begin{equation}
    \mathbb{C}_q^{t,p} := \{ \beta \in C_q^{t+p} | \partial_q^{t,p} (\beta) \in C_{q-1}^t \}.
\end{equation}
On this subset, one may define the $p$-persistent $q$-boundary operator denoted by $\hat{\partial}_q^{t,p}: \mathbb{C}_q^{t,p} \rightarrow C_{q-1}^t$ and corresponding adjoint operator $(\hat{\partial}_q^{t,p})^* : C_{q-1}^t \rightarrow  \mathbb{C}_q^{t,p}$, as before. The matrix representation of the $p$-persistent $q$-boundary operator in simplicial basis is then $\mathcal{B}_{q+1}^{t,p}$, and the matrix representation of the adjoint operator is again the transpose. This allows us to define the $q$-order $p$-persistent Laplacian matrix as: 
\begin{equation}
    \mathcal{L}_q^{t,p} := \mathcal{B}_{q+1}^{t,p}(\mathcal{B}_{q+1}^{t,p})^T + (\mathcal{B}_q^t)^T \mathcal{B}_q^t.
\end{equation}
We may again recognize the multiplicity of zero in the spectrum of $\mathcal{L}_q^{t,p}$ as the $q$'th order $p$-persistent Betti number $\beta_q^{t,p}$ which counts the number of (independent) $q$-dimensional voids in $K_t$ that still exists in $K_{t+p}$ \cite{wang2020persistent}. We can then see how the $q$'th-order Laplacian is actually just a special case of the $q$'th-order 0-persistent Laplacian at a simplicial complex $K_t$, or rather, at a snapshot of the filtration.

We can capture a more thorough view of the spatial features of our data by focusing on the 0-persistent Laplacian as we have done in previous works \cite{cottrell2023plpca}. Specifically, we calculate Vietoris-Rips complexes by varying a filtration parameter on the weighted entries of our Laplacian matrix, which correspond to the weighted edges in our graph structure. By gradually increasing a distance threshold, we induce a sequence of simplicial complexes and subgraphs to analyze. In previous works on Persistent Laplacian-enhanced PCA, we provided a convenient computational method for this. For a graph Laplcian matrix $L$, observe: 
\begin{equation}
L = (l_{ij}), l_{ij} =
    \begin{cases}
    l_{ij}, i \neq j, i,j = 1,...,n \\
    l_{ii} = -\sum_{j=1}^{n}l_{ij}.
    \end{cases} 
\end{equation}
For $i \neq j$, let $l_{\text{max}} = \text{max}(l_{ij})$, $l_{\text{min}} = \text{min}(l_{ij}), d = l_{\text{max}} - l_{\text{min}}$.
Set the $t^{th}$ Persistent Laplacian $L^t, t = 1,...,p$: 
\begin{equation}
L^t = (l_{ij}^t), l_{ij}^t = 
    \begin{cases}
    0, \text{ if } l_{ij} \leq (t/p)d + l_{\text{min}} \\
    -1, \text{otherwise}.
    \end{cases}
\end{equation}
Here, 
\begin{equation}
l_{ii}^t = -\sum_{j=1}^nl_{ij}^t.
\end{equation}
We then weight each $L^t$ in the sequence and sum to consolidate each subgraph into a single term, denoted $PL$. This new term should encode the persistence of topological features as the filtration progresses over multiple scales 
\begin{equation}\label{en:22}
PL := \sum_{t=1}^p\zeta_tL^t.
\end{equation} 
The optimal $\zeta$ weightings are hyper parameters which are obtained via Grid Search. Ideally, we should recognize which scales of connectivity contribute the most important information to our analysis, and place greater emphasis on that corresponding Laplacian matrix in the sum. More details regarding this procedure can be found in our previous works \cite{cottrell2023plpca}. Substituting this $PL$ term into Eq. \ref{eq:RgLSPCA} gives rise to Robust Persistent Laplacian Sparse PCA (RpLSPCA) which is obtained via the following optimization formula: \begin{align}\label{eq:RpLSPCA}
		\min_{U,Q}\|X - UQ^T\|_{2,1} + \beta\|Q\|_{2,1} + \gamma \text{Tr}(Q^T(PL)Q), \quad \text{s.t } Q^TQ = I_n.
	\end{align}
 For convenience, we can also refer to this method simply as Topological PCA (tPCA). This method should better retain geometrical structure information by emphasizing topological features that are persistent at multiple scales through the harmonic spectra, while the non harmonic spectra contributes other geometric shape information.
 
 \subsection{KNN Induced Persistent Laplacians}
\begin{figure}[H]
  \centering
    \includegraphics[width=0.8\linewidth]{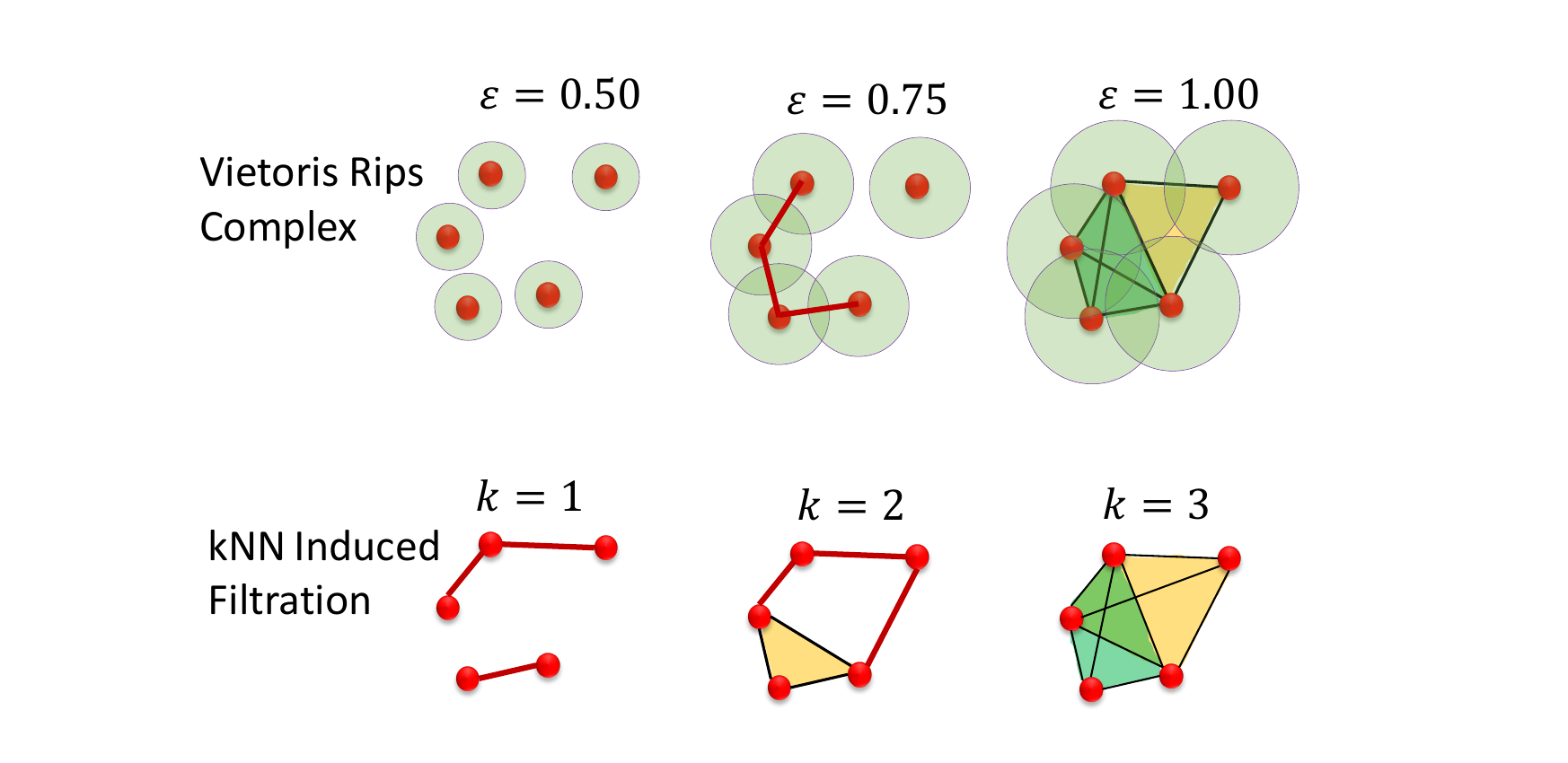}
  \caption{Comparison of inducing filtration via Vietoris Rips Complex, which is reliant on a chosen distance threshold $\epsilon$, and k-nearest-neighbors induced filtration, which is induced by varying the number of nearest neighbors at each node in our graph. }\label{fig:fil}
\end{figure}

 Rather than varying a distance threshold on the entries of our weighted Laplacian matrix to induce filtration, we can instead vary the number of neighbors we use to construct each graph structure. This construction has been used in the past to establish kNN-based persistent homology techniques, and extends naturally to the construction of a sequence of Persistent Laplacians\cite{le2022persistent}. Observe Figure \ref{fig:fil}. We then can replace  PLs  in  Eq. \ref{en:22} by kNN-PLs
 \begin{equation}
     L^k = (l_{ij}^k), l_{ij}^k = 
    \begin{cases}
    -1,\text{ if } i \neq j \text{ and }\mathbf{x}_j \in \mathcal{N}_k(\mathbf{x}_i) \\
    0, \text{ otherwise}
    \end{cases}
    \label{eq:18}
 \end{equation}

  We vary $k$ from 1 to $p$ to establish a sequence of $p$ Persistent Laplacians with different connectivities at each scale of the filtration. Again, diagonal entries are set equal to negative row sums, giving the total connectivity of each vertex. While the Vietoris-Rips based construction requires a choice of scale parameter for the Gaussian Kernel weighting of our graph Laplacian for the sake of normalization, the kNN construction eliminates the need for this since the filtrations do not depend on varying a distance threshold \cite{le2022persistent}. Furthermore, the sparseness of scRNA-seq data can affect distance measurements, leading to difficulties in establishing meaningful connections between cells when reliant on a distance metric. By inducing filtration through varying the number of neighbors rather than a distance threshold, the result is then a more standardized sequence of connectivity. We also tend to observe the formation of more connected cycles via this method, which lends itself to richer, more interesting topological features in our data. We now validate the performance of these new methods against other enhanced PCA procedures, as well as NMF and tSNE, on several benchmark scRNA-seq datasets.
	
	\section{Results}
	
		\subsection{Data and Preprocessing}
	\autoref{tab: dataset} shows the summary of the data, including the GEO accession ID, reference, source organism, number of samples, number of genes, and number of cell types. For each dataset, log-transform was applied, and genes with values less than $10^{-6}$ were set to zero. Afterward, $20\% - 25\%$ of the lowest variance genes were dropped. In certain instances where a class had fewer than 15 samples, we dropped that class from the analysis. For the PCA methods, values were then demeaned and scaled by the standard deviation, while for NMF, values were standardized by sklearn's NMF function. Note the differing dimensionality and number of cell types of each dataset, underscoring the comprehensiveness of our proposed methods. 
	\begin{table}[H]
		\centering
		\caption{Accession ID, source organism, and the counts for samples, genes, cell types and normalization for 11 datasets}
		\begin{tabular}{|c|c c c c  c  |} \hline
			Accession ID & Reference & Organism & Samples & Genes & Cell types  \\ \hline
			GSE67835 & Darmanis \cite{Darmanis2015ASO} & Human & 420 & 22084 & 8 \\
			GSE75748 cell & Chu \cite{article5} & Human & 1018 & 19097 & 7 \\
			GSE75748 time & Chu \cite{article5} & Human & 758 & 19189 & 6  \\
			GSE82187 & Gokce \cite{Gokce2016CellularTO} & Mouse & 705 & 18840 & 10 \\
			GSE94820 & Villani \cite{doi:10.1126/science.aah4573} & Human & 1140 & 26593 & 5 \\ 
            GSE84133human1 &Veres \cite{article6} &Human &1937 &20125 &14  \\
            GSE84133human2 &Veres \cite{article6} &Human &1724 &20125 &14   \\
            GSE84133human3 &Veres \cite{article6} &Human &3605 &20125 &14   \\ 
            GSE84133mouse1 &Veres \cite{article6} &Mouse &822 &14878 &13   \\
            GSE84133mouse2 &Veres \cite{article6} &Mouse &1064 &14878 &13   \\ 
            GSE45719
            &Deng \cite{doi:10.1126/science.1245316} &Mouse &300 &22431 &8  \\ \hline
            
		\end{tabular}
		\label{tab: dataset}
	\end{table}

	\subsection{Evaluation Metrics}
	\subsubsection{Adjusted Rand Index (ARI)}
	ARI describes how well two clusterings agree with each other by comparing pairs of data points and their respective class assignments. It also can account for possible random agreement and adjusts the similarity score accordingly, assigning a value in the range of -1 to 1. A value of 1 indicates a perfect agreement between clusterings, a value of 0 indicates a random chance agreement, and a value of -1 suggests that the clusterings are less similar than they would be by chance. For two clusterings $X = \{ X_1,...,X_r\}$ and $Y=\{Y_1,...,Y_s\}$, we construct a contingency table $A \in \mathbb{R}^{r \times s}$ with elements $a_{ij}$ which describe the overlap between $X_i$ and $Y_j$. We then take row sums and column sums to obtain another set of values: $\{ q_1,...,q_r\}$ is the set of row sums and $\{p_1,...,p_s \}$ is the set of column sums. We can then define the Adjusted Rand Index as: 
 
 \begin{equation}
     \text{ARI} = \frac{\sum_{ij} \binom{a_{ij}}{2} - (\sum_i \binom{q_i}{2}\sum_j \binom{p_j}{2})} {\frac{1}{2} (\sum_i \binom{q_i}{2} + \sum_j \binom{p_j}{2}) - (\sum_i \binom{q_i}{2}\sum_j \binom{p_j}{2})}
 \end{equation}
 
	\subsubsection{Normalized Mutual Information (NMI)}
	Mutual Information considers a split of the data according to clusters and a split according to true class labels, and measures how these splittings agree with each other. NMI then corrects for any bias and normalizes the scores between 0 and 1. A value of 1 indicates a perfect agreement between the splittings while a value of 0 indicates random chance agreement. The mathematical definition of NMI is given as: 
 \begin{equation}
     \text{NMI}(T,P) = \frac{\text{MI}(T,P)}{(\text{E}(T) +\text{E}(P))/2 }
 \end{equation}
 Where MI($ \cdot,\cdot $) and E($\cdot $) represent mutual information and entropy and $T,P$ represent the true and predicted cluster labels, respectively. 

    \subsubsection{Classification Metrics}

Regarding the evaluation metrics used to measure performance for classification tasks, beyond simple accuracy, there are several metrics that are commonly considered. Notably, Precision, Recall, and F1-Score. Below we list the mathematical definition of each. 

\begin{align}
    \text{Recall} &= \frac{\text{True Positive}} { \text{True Positive + False Negative} }\\
    \text{Precision} &= \frac{\text{True Positive}} {\text{True Positive + False Positive}} \\
    \text{F1-Score} &= 2 \frac {\text{Precision} \times \text{Recall}}{\text{Precision}+\text{Recall}}
\end{align}

 We note that the F1-score is particularly relevant as it accounts for both precision and recall, making it more robust to class imbalances within the data. In our case, there are noticeable imbalances between cell types in each of the tested dataset, so we emphasize this metric as the most informative in measuring performance.

Given that our datasets generally contain multiple classes, the evaluation criterion we employ is the mean for each cell type. This evaluation approach is commonly known as a macro metric, where performance measures are calculated for each cell type individually and then averaged to obtain an overall score.

\begin{align}
    \text{Macro-Recall} &= \frac{1}{c}   \sum_{i=1}^{c} \text{Recall}_i    \\
    \text{Macro-Precision} &= \frac{1}{c}  \sum_{i=1}^{c} \text{Precision}_i   \\
    \text{Macro-F1} &= 2   \frac{\text{Macro-Precision} \times \text{Macro-Recall}}{\text{Macro-Precision}+\text{Macro-Recall}}
\end{align}

\subsubsection{Residue Similarity Scores}
To enhance visualization, Residue Similarity (RS) scores can be computed \cite{hozumi2022ccp}. Traditional visualization techniques often involve reducing the data to two or three dimensions, which may result in the loss of structure and integrity in multiclass data. R-S plots were introduced as a method to visualize results while better preserving the underlying structure of the data.

An R-S plot consists of two main components: the residue score and the similarity score. The residue score is calculated as the sum of distances between classes, capturing the dissimilarity between them. On the other hand, the similarity score represents the average similarity within each class, indicating the degree of similarity between instances belonging to the same class. By considering both scores, R-S plots provide a comprehensive representation of the data's structure in a visualization.

Given data of the form $ \{(\vec{x}_i, y_i) | \vec{x}_i \in \mathbb{R}^N, y_i \in \mathbb{Z}_l \}_{i=1}^M$, we have $y_i$ representing the class label of our $i$th data point $\vec{x}_i \in \mathbf{X}$. Say that our data has $N$ samples, $M$ features, and $L$ classes. We can then partition our dataset $\mathbf{X}$ into subsets containing each of the classes by taking $\mathcal{C}_l = \{ \vec{x}_i \in \mathbf{X} | y_i = l\}$. For each class $l$ we then define the residue score as follows: 
\begin{equation}
    R_i := R(\vec{x}_i) = \frac{1}{R_{\rm max}} \sum_{\vec{x}_j \notin \mathcal{C}_l} \lVert \vec{x}_i - \vec{x}_j
 \rVert,
\end{equation} 
where $\lVert \cdot \rVert$ denotes the Euclidean distance between vectors and $R_{\rm max}$ is the maximal residue score for that subset. The similarity score, meanwhile, is given as: 

\begin{equation}
    S_i := S(\vec{x}_i) = \frac{1}{\lvert \mathcal{C}_l \rvert} \sum_{\vec{x}_j \in \mathcal{C}_l}\left(1 - \frac{\lVert \vec{x}_i - \vec{x}_j \rVert} {d_{\rm max}}\right), 
\end{equation} 

where $d_{\rm max}$ is the maximal pairwise distance of the dataset. For constructing R-S plots, we then take $R(\vec{x})$ to be the ${x}$-axis and $S(\vec{x})$ to be the $ {y}$-axis. 
 
    \subsection{Comparison of tPCA and Other Methods for KMeans Clustering} 
	
	We tested tPCA's performance on the datasets described in \autoref{tab: dataset}, and compared it to PCA, sPCA, RgLSPCA,  NMF, UMAP, and  tSNE. For this analysis, we performed K-Means clustering after reducing the dimensionality of each dataset such that the number of dimensions equals the number of clusters. To ensure the greatest accuracy in our analysis, we used sklearn's KMeans function, and increased the number of times the k-means algorithm is run with different centroid seeds from 10 to 150. We then performed the clustering with 30 random instances and considered the average performance. To further facilitate a fair and accurate comparison, we utilized sklearn's NMF function, initialized as non-negative random matrices with values scaled by the square root of the mean of $X$ and divided by the number of components. We also increased the number of maximum iterations from 200 to 300, and took the average performance over 20 random initializations. Likewise, we also compare our method to sklearn's tSNE over 20 random intializations with maximum iterations increased to 300. For UMAP, we used the default minimum distance allowed for packing points together in the embedding space, the low value generally should improve clustering performance by providing a cleaner separation between clusters. We used the default number of nearest neighbors in UMAP's kNN structure, which was 15. This value is the same as the initial number of neighbors used to describe the manifold structure in tPCA. 
 
 We do, however, note certain weaknesses with NMF, UMAP, and tSNE in this analysis. First, methods such as NMF tend to perform poorly when reducing to such low dimensionality, while methods such as tSNE and UMAP have notable issues with structure preservation. Specifically, neither method is capable of strictly preserving the density or distance in our data. Therefore, we expect that our procedure should prove substantially more effective as a preprocessing step than any of these methods for clustering via KMeans or classifying via K-Nearest-Neighbors, as both models are reliant on the intrinsic distance and density in the data. For the clustering analysis, a summary of our results for Adjusted Rand Index and Normalized Mutual Information can be found in Tables \ref{tab: comparison RPL vs Rest},\ref{tab: comparison RPL vs Rest2}. For kNN-Induced tPCA, when results are shown as $(\cdot)*$, this implies that a generic connectivity weighting was used, and there was no parameter optimization needed on that dataset to obtain nearly optimal performance.  

\begin{table}[H]
    \centering
    \caption{Comparison of tPCA and other methods for Adjusted Rand Index}
    \label{tab: comparison RPL vs Rest}
    {
    \begin{tabular}{ c|cccccccccc } 
    \toprule
    Dataset and Method & kNN-tPCA & tPCA & RgLSPCA  & sPCA & PCA & NMF & tSNE & UMAP \\
    \midrule 
    \multirow{1}{8em}{GSE67835} &(0.9496)* &\textbf{0.9552} &0.9496 &0.7625 &0.7604 &0.6926 & 0.4200  & 0.5579  \\
    \multirow{1}{8em}{GSE75748cell}&0.7475  &\textbf{0.7789} &0.7454 &0.7440 &0.7440 &0.7530 &0.5301 & 0.6511 \\
    \multirow{1}{8em}{GSE75748time}& (0.7123)* &\textbf{0.7129} &0.7036  &0.6128 &0.6128 &0.6852 & 0.3746 & 0.3677\\
    \multirow{1}{8em}{GSE82187} & \textbf{(0.9932)*} &\textbf{0.9932} &0.7622  &0.7530 &0.7530 &0.5113 & 0.4727 & 0.5193\\
    \multirow{1}{8em}{GSE94820}&0.5731  &\textbf{0.5749}  &0.5159 & 0.5396&0.5396 &0.5400 &0.3820 & 0.3917\\
    \multirow{1}{8em}{GSE84133human1}&(0.7942)* &\textbf{0.8065} &0.7914  &0.5675 &0.5675 &0.6436 &0.4446 & 0.5396\\
    \multirow{1}{8em}{GSE84133human2}& \textbf{(0.9277)*} &\textbf{0.9277} &0.9162  &0.8090 &0.8090 &0.6168 &0.6237 & 0.5680\\
    \multirow{1}{8em}{GSE84133human3}& \textbf{(0.8727)*} &0.8722 &0.7799  &0.6976 &0.6976 &0.6061 &0.5964 & 0.7016\\
    \multirow{1}{8em}{GSE84133mouse1} &\textbf{0.7746}  &0.7699 &0.7699  &0.7688 &0.6320 &0.5417  &0.5391& 0.4293\\
    \multirow{1}{8em}{GSE84133mouse2}&\textbf{0.6172} &0.6139 &0.6165 &0.5041 &0.5041 &0.4029 &0.3978& 0.3851 \\
    \multirow{1}{8em}{GSE45719}&0.4262 &\textbf{0.4386} &0.4212  &0.4133 &0.4133 &0.4073 &0.4068 & 0.4088\\
    \bottomrule 
    \end{tabular} 
    }
\end{table}
We see from the results in Table \ref{tab: comparison RPL vs Rest} that persistent Laplacian-enhanced PCA is able to outperform not only the other enhanced PCA procedures, but also NMF, UMAP, and tSNE, on all of the tested datasets for Adjusted Rand Index. We specifically note the superior performance on GSE82187. our tPCA outperformed RgLSPCA's ARI score by a considerable margin of 0.231. These results clearly indicate the superior ability to preserve local non-linear geometric structure achieved by persistent Laplacian regularization compared to graph Laplacian regularization, and the result is superior performance in clustering analysis. Furthermore, in several instances KNN-Induced Laplacian regularization was able to match the performance of the standard construction without any optimization. Overall, it was shown to at least outperform the other procedures on all but one tested dataset, and with a fraction of the effort required for parameter search. 

\begin{table}[H]
    \centering
    \caption{Comparison of tPCA and other methods for Normalized Mutual Information}
    \label{tab: comparison RPL vs Rest2}
    {
    \begin{tabular}{ c|cccccccccc } 
    \toprule
    Dataset and Method & kNN-tPCA & tPCA & RgLSPCA  & sPCA & PCA & NMF  & tSNE & UMAP\\
    \midrule 
    \multirow{1}{8em}{GSE67835}& (0.9224)*&\textbf{0.9275} &0.9224  &0.8192 &0.8174 &0.78630&0.5817  & 0.7272   \\
    \multirow{1}{8em}{GSE75748cell}&0.8850 &\textbf{0.9230} &0.8825  &0.8810 &0.8810 &0.9036 &0.6940 & 0.7956 \\
    \multirow{1}{8em}{GSE75748time}&(0.8276)* &\textbf{0.8338} &0.8248  &0.7220 &0.7220 &0.8074 & 0.4847 & 0.5047\\
    \multirow{1}{8em}{GSE82187}& \textbf{(0.9853)*} &\textbf{0.9853} &0.8967  &0.8995 &0.8995 &0.8040  & 0.6849 & 0.7841\\
    \multirow{1}{8em}{GSE94820}&\textbf{0.6757} &0.6738 &0.6275  &0.6367 &0.6367 &0.6444  &0.4949 & 0.4831 \\
    \multirow{1}{8em}{GSE84133human1}&(0.8412)* &\textbf{0.8604} &0.8391  &0.7905 &0.7905 &0.8142 &0.6550 & 0.7748\\
    \multirow{1}{8em}{GSE84133human2}& \textbf{(0.9158)*} &\textbf{0.9158} &0.9036  &0.7816 &0.7816 &0.7603  & 0.7044 & 0.7571\\
    \multirow{1}{8em}{GSE84133human3} & \textbf{(0.8742)*} &0.8736 &0.8131  &0.8159 &0.8160 &0.7644  &0.6921 & 0.8226\\
    \multirow{1}{8em}{GSE84133mouse1}&\textbf{0.8587} &0.8514 &0.8581 &0.8473 &0.7480 &0.6847 &0.6159& 0.6671  \\
    \multirow{1}{8em}{GSE84133mouse2}&\textbf{0.7924}  &0.7918 &0.7923  &0.7094 &0.7095 &0.6424  &0.5860 & 0.6403\\
    \multirow{1}{8em}{GSE45719}&0.6549 &\textbf{0.6747} &0.6116  &0.6022 &0.6022 &0.5897  &0.5960 & 0.6034\\
    \bottomrule 
    \end{tabular} 
    }
\end{table} 

Once again, the results in Table \ref{tab: comparison RPL vs Rest2} showcase the superiority of tPCA in all tested cases for NMI. We again note the remarkably superior performance of our method on the GSE82187 dataset specifically, where we outperform RgLSPCA in NMI by 0.089. Again, we also note the ability of the kNN-tPCA to provide optimal or near optimal performance compared to the distance based construction while not requiring an extensive parameter tuning procedure. Now, we can average the performance in each metric over all tested datasets to reveal the extent to which tPCA and kNN-tPCA outperform the other methods overall, across our 11 tested datasets.

\begin{table}[H]
    \centering
    \caption{Comparison of tPCA and other methods for each performance metric averaged over all tested datasets}
    \label{tab: avg}
    {
    \begin{tabular}{ lcc } 
    \toprule
    Method & ARI & NMI \\
    \midrule 
    \multirow{1}{8em}{kNN-tPCA}   & \textbf{0.7625} & \textbf{0.8393} \\
    \multirow{1}{8em}{tPCA}   & \textbf{0.7676} & \textbf{0.8464} \\
    \multirow{1}{8em}{RgLSPCA}   & \text{0.7247} & \text{0.8156} \\
    \multirow{1}{8em}{sPCA}   & \text{0.6520} & \text{0.7732} \\
    \multirow{1}{8em}{PCA}   & \text{0.6401} & \text{0.7640} \\
     \multirow{1}{8em}{NMF}   & \text{0.5818} & \text{0.7455}\\
     \multirow{1}{8em}{tSNE}   & 0.4716 & 0.6172 \\
     \multirow{1}{8em}{UMAP}   & 0.5018 & 0.6872 \\
    \bottomrule 
    \end{tabular} 
    }
    
\end{table}

\begin{figure}[H]
  \centering
    \includegraphics[width=0.7\linewidth]{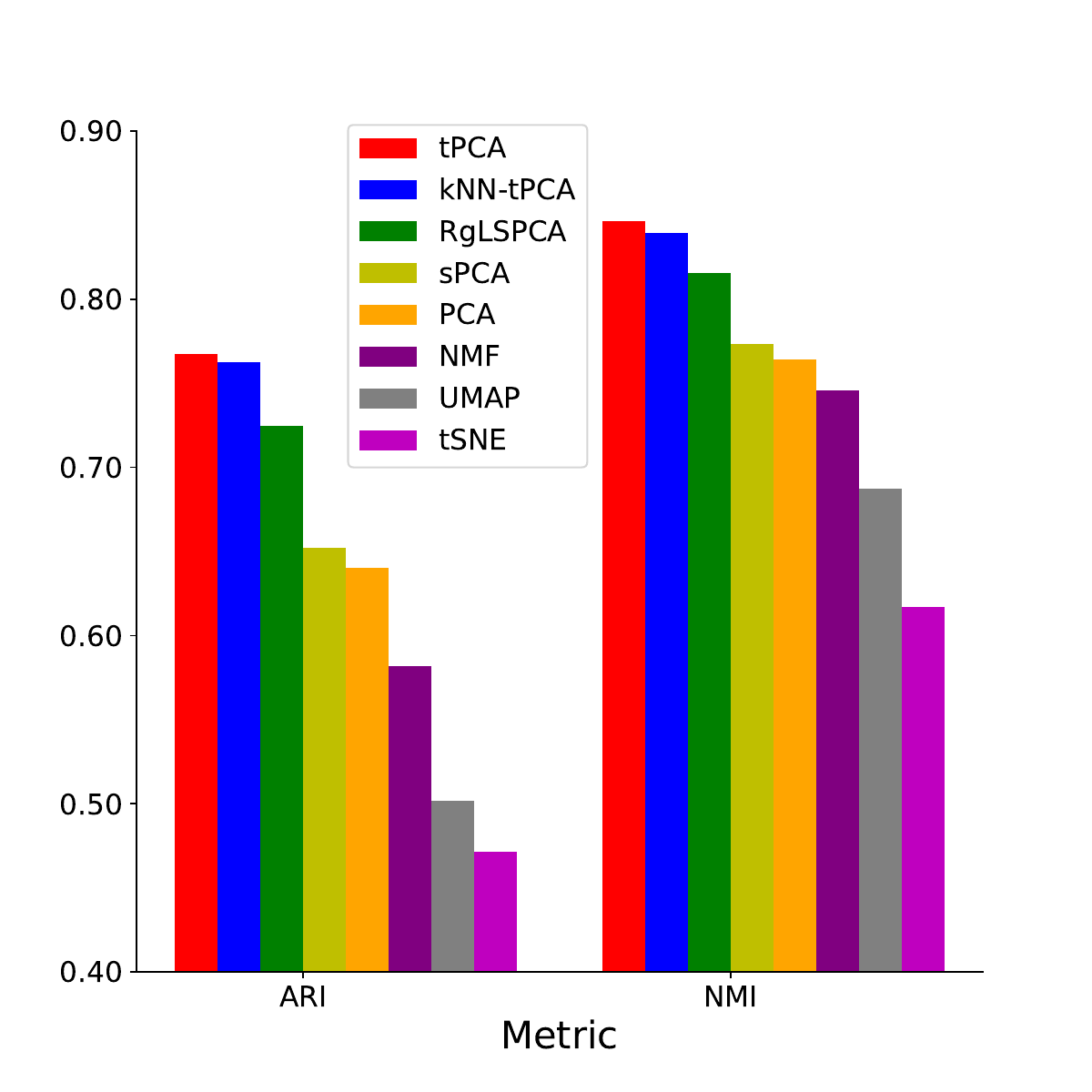}
  \caption{NMF and ARI comparisons for each method averaged over all datasets.}\label{fig:compfig}
\end{figure}

We see from the final results in Table \ref{tab: avg} that, on average, tPCA outperforms NMF by a significant measure of 31.92\% for ARI and 13.53\% for NMI, and RgLSPCA by 3.78\% for NMI and 5.92\% for ARI. kNN-tPCA, meanwhile, outperforms NMF by 31.05\% for ARI and 12.58\% for NMI, and RgLSPCA by 2.91\% for NMI and 5.22\% for ARI. To intuitively illustrate this point, in Figure \ref{fig:compfig} we provide a barplot comparing the performance metrics of the mentioned procedures averaged over each of the 11 tested datasets.

 From the depicted image it is clearly evident that both methods for tPCA are superior to all other tested dimensionality reduction techniques, particularly other PCA enhancements. We especially emphasize the superiority of kNN-tPCA given the significantly reduced need for parameter optimization with this method. These results strongly reaffirms the importance of incorporating the topological information and multi-scale analysis that is possible with topological PCA into a dimensionality reduction technique. While previous techniques can capture some geometrical structure information through graph Laplacian regularization, we see that incorporating the additional filtrations greatly improves performance. Having confirmed the efficacy of our methods, we can move to examining the impact that kNN-induced filtration has on the scale of parameter tuning in more detail, as well as comparing different visualization techniques of the Eigen-Genes each method produces.

 \subsection{Comparison of kNN-tPCA and Other Methods for   Classification} 

To further validate the efficacy of our proposed method, we can supplement these clustering results with a   classification study using kNN. The classification of various cell types begins by randomly splitting our gene expression data into training and testing sets. The kNN model is trained on 60\% of the data, and then tested on the remaining 40\%. To mitigate the impact of data distribution, we employed a 5-fold cross-validation approach. The classification accuracy was calculated as the average performance over five repetitions. The mean accuracy of the classification was then recorded for subspace dimensions ranging from \{100, 90, ..., 10, 1\}. The results of this analysis can be seen in Tables \ref{tab:knn-tab-1} and  \ref{tab:knn-tab-2}. 

 {\small
\begin{table}[H]
    \centering
    \setlength\tabcolsep{4pt}
    \captionsetup{margin=0.1cm}
    \caption{Comparison of average results for kNN-tPCA and Other Methods for   Classification after dimensionality reduction to $m = 1,10,...,100$}
    \label{tab:knn-tab-1}
    \begin{tabular}{ c|lccccc } 
    \toprule
    Dataset & Method & Mean-ACC & Mean Macro-REC & Mean Macro-PRE & Mean Macro-F1 \\
    \midrule
    GSE67835 & \textbf{kNN-tPCA} &\textbf{0.8909} &\textbf{0.8345} &\textbf{0.8672} &\textbf{0.8455} \\
            & RgLSPCA &0.8808 &0.8032 &0.8358 &0.8084\\
             & sPCA &0.8319 &0.6872 &0.7918 &0.7052 \\
             & PCA &0.7732 &0.6172 &0.7631 &0.63805 \\
             & NMF &0.4136 &0.2749 &0.2740 &0.2620 \\
             & tSNE &0.6470 &0.4986&0.5037 &0.4835 \\
             & UMAP &0.1793&0.1380 &0.0509 &0.0588 \\
    \midrule
    GSE75748cell & \textbf{kNN-tPCA} &\textbf{0.9568} &\textbf{0.9267} &\textbf{0.9332} &\textbf{0.9291} \\
            & RgLSPCA &0.9499 &0.9204 &0.9260 &0.9218\\
             & sPCA &0.9305 &0.9006 &0.9175 &0.9046 \\
             & PCA &0.7222 &0.5731 &0.6603 &0.5736 \\
             & NMF &0.4210 &0.3784 &0.3784 &0.3784 \\
             & tSNE &0.5292 &0.5257&0.5402 &0.5224 \\
             & UMAP &0.3505& 0.3554&0.2485 &0.2678 \\
    \midrule
    GSE75748time & \textbf{kNN-tPCA} &\textbf{0.8222} &\textbf{0.8006} &\textbf{0.8874} &\textbf{0.8068} \\
            & RgLSPCA &0.7928 &0.7660 &0.8692 &0.7667\\
             & sPCA &0.7590 &0.7307 &0.8353 &0.7222 \\
             & PCA &0.7587 &0.7303 &0.8352 &0.7219 \\
             & NMF &0.3792 &0.3501 &0.3625 &0.3312 \\
             & tSNE &0.3858 &0.3394&0.3309 &0.3178 \\
             & UMAP &0.2305&0.1975 &0.1006 &0.1114 \\
    \midrule
    GSE82187 & \textbf{kNN-tPCA} &\textbf{0.9028} &\textbf{0.8520} &\textbf{0.9115} &\textbf{0.8710} \\
            & RgLSPCA &0.8422 &0.7280 &0.8273 &0.7489\\
             & sPCA &0.7357 &0.5917 &0.6890 &0.5958 \\
             & PCA &0.7222 &0.5731 &0.6603 &0.5736 \\
             & NMF &0.6164 &0.3831 &0.4070 &0.3773 \\
             & tSNE &0.5887 &0.5648&0.5710 &0.5621 \\
             & UMAP &0.4791&0.1896 &0.1272 &0.1374 \\
    \midrule
    GSE94820 & \textbf{kNN-tPCA} &\textbf{0.8914} &\textbf{0.8346} &\textbf{0.8677} &\textbf{0.8455} \\
            & RgLSPCA &0.8803 &0.8029 &0.8349 &0.8072\\
             & sPCA &0.8319 &0.6872 &0.7918 &0.7052 \\
             & PCA &0.7732 &0.6172 &0.7631 &0.6380 \\
             & NMF &0.6618 &0.4001 &0.4592 &0.4263 \\
             & tSNE &0.3330 &0.3288&0.3358 &0.3110 \\
             & UMAP &0.2485&0.1983 &0.0748 &0.0906 \\
    \midrule
    \end{tabular} 
\end{table}
}
We see from this first round of results that kNN-tPCA provides a stellar improvement to performance metrics for   classifications using kNN, carried out after dimensionality reduction. Notably, we observe a 2.95\% improvement in F1-Score when compared to the standard graph regularization in tPCA and a remarkable 17.5\% improvement when compared to traditional PCA. Compared to other dimensionality reduction techniques such as UMAP, tSNE, and NMF, the results are even more significant. This demonstrates the comprehensiveness of tPCA in being able to reduce data to a variety of embeddding dimensions while also preserving important structural information in the data. The results in Table \ref{tab:knn-tab-1} as well as those in Table \ref{tab:knn-tab-2} demonstrate that kNN induced Persistent Laplacian regularization for PCA is a comprehensive method capable of enhancing the performance of classification tasks for Single Cell RNA-Sequence data analysis. Combined with the results in Table \ref{tab: avg}, we can conclude that tPCA is a superior dimensionality reduction technique for a variety of Machine Learning methods. 

\begin{center}
    \begin{figure}[H]
        \centering
        \includegraphics[width=\linewidth]{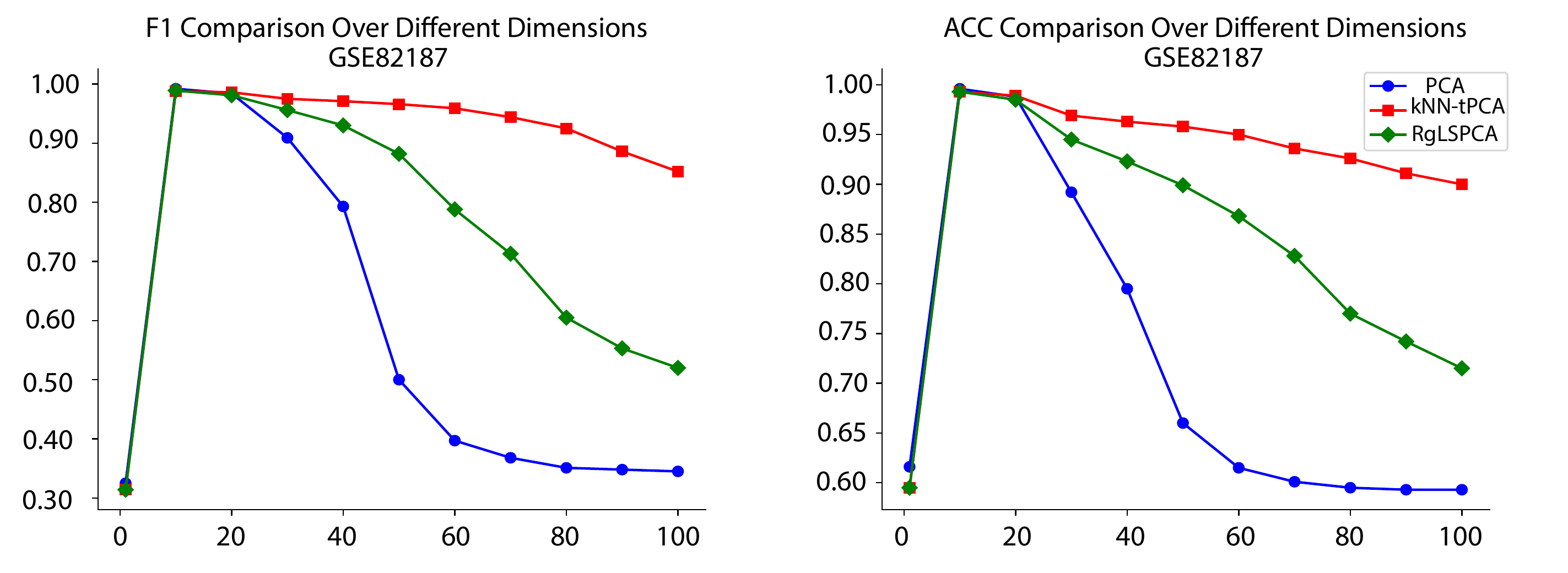}
        \caption{Distributions of ACC and F1 performance for PCA, RgLSPCA, and kNN-tPCA as we vary the number of reduced subspace dimensions.}
        \label{fig:acc_plots}
    \end{figure}
\end{center}

To intuitively illustrate these results, in Figure \ref{fig:acc_plots} we have provided an illustration depicting the distribution of Accuracy and F1 performance between PCA, RgLSPCA, and kNN-tPCA as we vary the dimensionality of our reduced space on GSE82187. This clearly indicates the superior performance of our proposed method across a wide range of reduced dimensions, especially as the number of dimensions grows larger, where PCA typically suffers from stability issues. This clearly further validates our findings. 

{\small
\begin{table}[H]
    \centering
    \setlength\tabcolsep{4pt}
    \captionsetup{margin=0.1cm}
    \caption{Comparison of average results for kNN-tPCA and Other Methods for kNN Classification after dimensionality reduction to $m = 1,10,...,100$}
    \label{tab:knn-tab-2}
    \begin{tabular}{ c|lccccc } 
    \toprule
    Dataset & Method & Mean-ACC & Mean Macro-REC & Mean Macro-PRE & Mean Macro-F1 \\
    \midrule
    GSE45719 & \textbf{kNN-tPCA} &\textbf{0.9282} &\textbf{0.9022} &\textbf{0.9119} &\textbf{0.9055} \\
            & RgLSPCA &0.9282 &0.9003 &0.9112 &0.9038\\
             & sPCA &0.8371 &0.8209 &0.8735 &0.8296 \\
             & PCA &0.8426 &0.8260 &0.8757 &0.8356 \\
             & NMF &0.2470 &0.2413 &0.2460 &0.2196 \\
             & tSNE &0.4376 &0.4452& 0.4426& 0.4302\\
             & UMAP & 0.1581&0.1454 &0.071 &0.0856 \\
    \midrule
    GSE84133human1 & \textbf{kNN-tPCA} &\textbf{0.8911} &\textbf{0.8209} &\textbf{0.8745} &\textbf{0.8417} \\
            & RgLSPCA &0.8837 &0.8075 &0.8718 &0.8308\\
             & sPCA &0.8279 &0.7679 &0.8633 &0.7951 \\
             & PCA &0.8279 &0.7680 &0.8633 &0.7952 \\
             & NMF &0.5109 &0.3973 &0.3800 &0.3672 \\
             & tSNE &0.7697 &0.6175&0.6180 &0.5960 \\
             & UMAP &0.1858 &0.1818 &0.0827 &0.0979 \\
    \midrule
    GSE84133human2 & \textbf{kNN-tPCA} &\textbf{0.9216} &\textbf{0.8761} &\textbf{0.8997} &\textbf{0.8829} \\
            & RgLSPCA &0.9157 &0.8687 &0.8993 &0.8765\\
             & sPCA &0.9178 &0.8657 &0.8861 &0.8727 \\
             & PCA &0.8973 &0.8271 &0.8903 &0.8395 \\
             & NMF &0.5383 &0.3652 &0.3718 &0.3521 \\
             & tSNE &0.5786 &0.5302&0.5651 &0.5258 \\
             & UMAP &0.1322&0.1759 &0.0758 &0.0903 \\
    \midrule
    GSE84133human3 & \textbf{kNN-tPCA} &\textbf{0.9062} &\textbf{0.8487} &\textbf{0.8758} &\textbf{0.8600} \\
            & RgLSPCA &0.9034 &0.8461 &0.8734 &0.8573\\
             & sPCA &0.8838 &0.8178 &0.8615 &0.8358 \\
             & PCA &0.8279 &0.7680 &0.8633 &0.7952 \\
             & NMF &0.5712 &0.4298 &0.4568 &0.3990 \\
             & tSNE &0.7226 &0.5889& 0.6358&0.5867 \\
             & UMAP &0.2047 &0.1773 &0.0948 &0.1054 \\
    \midrule
    GSE84133mouse1 & \textbf{kNN-tPCA} &\textbf{0.9172} &\textbf{0.8825} &\textbf{0.8995} &\textbf{0.8898} \\
            & RgLSPCA &0.9149 &0.8766 &0.8986 &0.8857\\
             & sPCA &0.9011 &0.8541 &0.8887 &0.8665 \\
             & PCA &0.9011 &0.8544 &0.8889 &0.8666 \\
             & NMF &0.5620 &0.4004 &0.4215 &0.3886 \\
             & tSNE &0.8928 &0.6995& 0.7092&0.6924 \\
             & UMAP &0.3541&0.2593 &0.1458 &0.1691 \\
    \midrule
    GSE84133mouse2 & \textbf{kNN-tPCA} &\textbf{0.9213} &\textbf{0.8963} &\textbf{0.9002} &\textbf{0.8976} \\
            & RgLSPCA &0.9209 &0.8953 &0.8995 &0.8968\\
             & sPCA &0.9018 &0.8542 &0.8894 &0.8667 \\
             & PCA &0.9011 &0.8544 &0.8889 &0.8666 \\
             & NMF &0.5403 &0.3428 &0.3567 &0.3309 \\
             & tSNE &0.6695 &0.3759&0.3752 &0.3627 \\
             & UMAP &0.2501&0.1563 &0.0897 &0.1008 \\
    \midrule
    \end{tabular} 
\end{table}
}

\begin{figure}[H]
  \centering
    \includegraphics[width=0.8\linewidth]{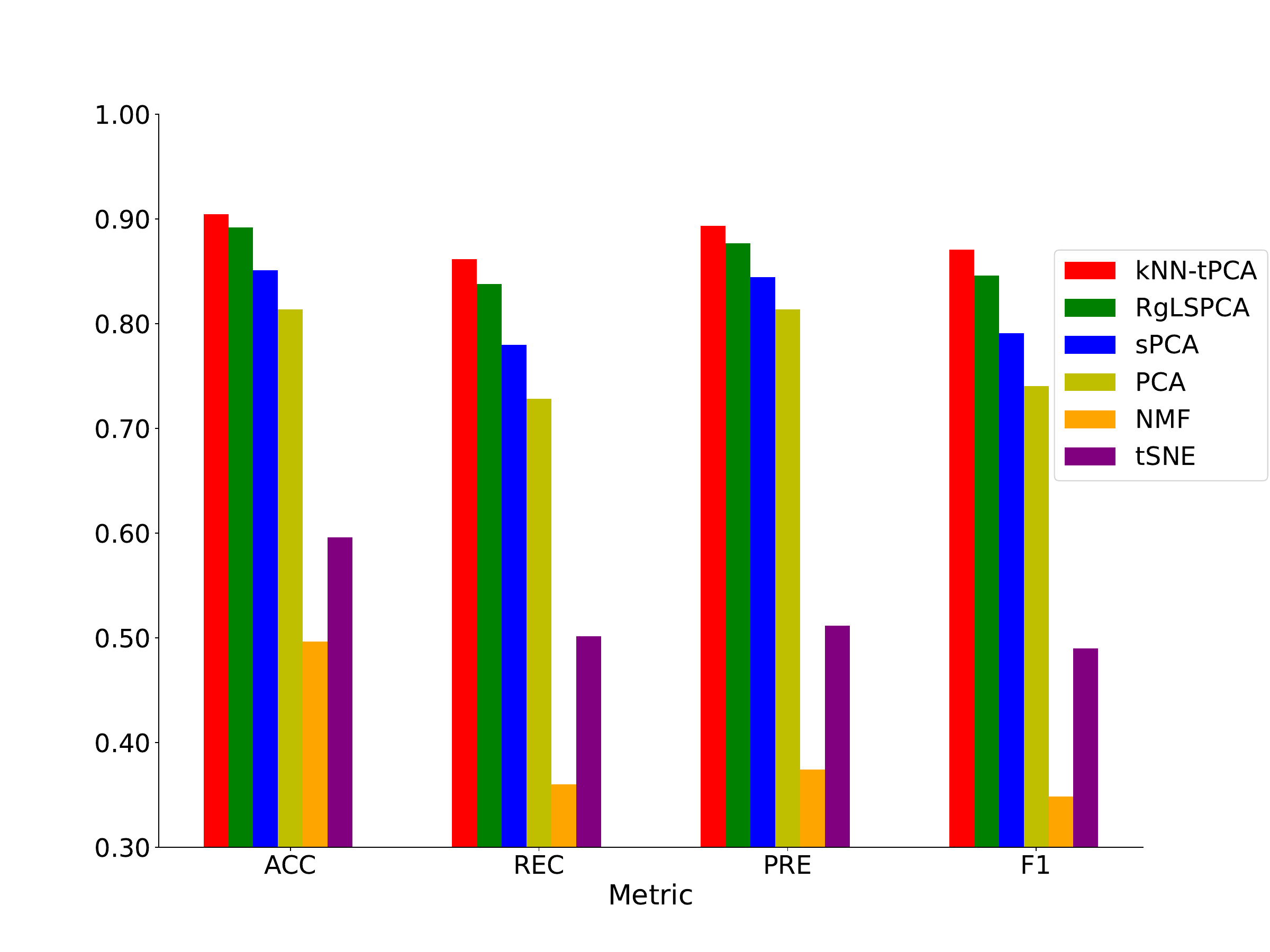}
  \caption{ACC, PRE, REC, and F1 comparisons for each methods averaged over all datasets.}\label{fig:avg2}
\end{figure}

 When we average each of the performance metrics over the 11 tested datasets, we can assess the total improvement that our method provides compared to other techniques as  shown in Table \ref{tab: avg2}. We can then intuitively visualize these results by examining Figure \ref{fig:avg2}, which clearly showcases the superiority of kNN-tPCA for classification tasks on a variety of datasets with different dimensionalities and data imbalances. 

\begin{table}[H]
    \centering
    \caption{Comparison of kNN-tPCA and other methods for each performance metric averaged over all tested datasets}
    \label{tab: avg2}
    {
    \begin{tabular}{ lcccc } 
    \toprule
    Method & ACC & REC & PRE & F1 \\
    \midrule 
    \multirow{1}{8em}{\textbf{kNN-tPCA}}   &\textbf{0.9045} &\textbf{0.8613} &\textbf{0.8935} &\textbf{0.8704} \\
    \multirow{1}{8em}{RgLSPCA}   &0.8920 &0.8377 &0.8770 &0.8458 \\
    \multirow{1}{8em}{sPCA}   &0.8507 &0.7798 &0.8443 &0.7908 \\
    \multirow{1}{8em}{PCA}   &0.8134 &0.7280 &0.8138 &0.7403 \\
     \multirow{1}{8em}{NMF}   &0.4965 &0.3603 &0.3739 &0.3484 \\
     \multirow{1}{8em}{tSNE}   &0.5958 &0.5013 &0.5115 &0.4900 \\
     \multirow{1}{8em}{UMAP}   &0.2520 &0.1977 &0.1056 &0.1195 \\
    \bottomrule 
    \end{tabular} 
    }
\end{table}

 We specifically note that, on average, kNN-tPCA outperforms RgLSPCA by a margin of 1.39\% for Macro-ACC, 1.88\% for Macro-PRE, 2.82\% for Macro-REC, and 2.91\% for Macro-F1. This clearly demonstrates the benefits of
incorporating multi-scale analysis through the inclusion of persistent Laplacians. Furthermore, we note that
our method outperforms traditional PCA by up to a considerable 18.3\% for
these metrics over the 11 datasets.

\section{Discussion}
\subsection{Parameter Analysis}
Regarding the optimization of hyper-parameters for tPCA, we especially note the presence of the $\zeta$ weights in the $PL$ term: 
  \begin{equation}
      PL := \sum_{t=1}^p\zeta_tL^t
  \end{equation}

  Which must be manually chosen for each dataset depending on the connectivity information that is most important. Specifically, for $p$ filtrations we generally consider a distribution of $\{1,1/2,...,1/p,0\}$ and perform a parameter search over this distribution, while also simultaneously searching for an optimal $\gamma$ value. However, for larger $p$ this clearly becomes an extremely computationally intensive task, with the number of parameter combinations equaling $((p+1)^p)(\text{Size of } \gamma \text{ distribution})$. Therefore, rather than considering all parameter combinations over this distribution at once, we can instead consider different combinations of scales of connectivity, say, long, middle, and close range. In other words, for 7 filtrations, testing combinations of $p=7,5,3$, and from there recognizing which scales contribute the most valuable information to narrow our search. Doing so reduces the number of combinations from $((p+1)^p)(\text{Size of } \gamma \text{ distribution})$ to $(m)((p+1)^3)(\text{Size of } \gamma \text{ distribution})$, where m is the number of connectivity combinations we need to test to achieve the best results. In practice, we found that generally $m =3$ obtained allowed us to obtain optimal performance, which is a considerable improvement from the traditional approach to grid search, though clearly still not preferable for practical purposes. 
  
\begin{figure}[H]
  \centering
    \includegraphics[width=0.6\linewidth]{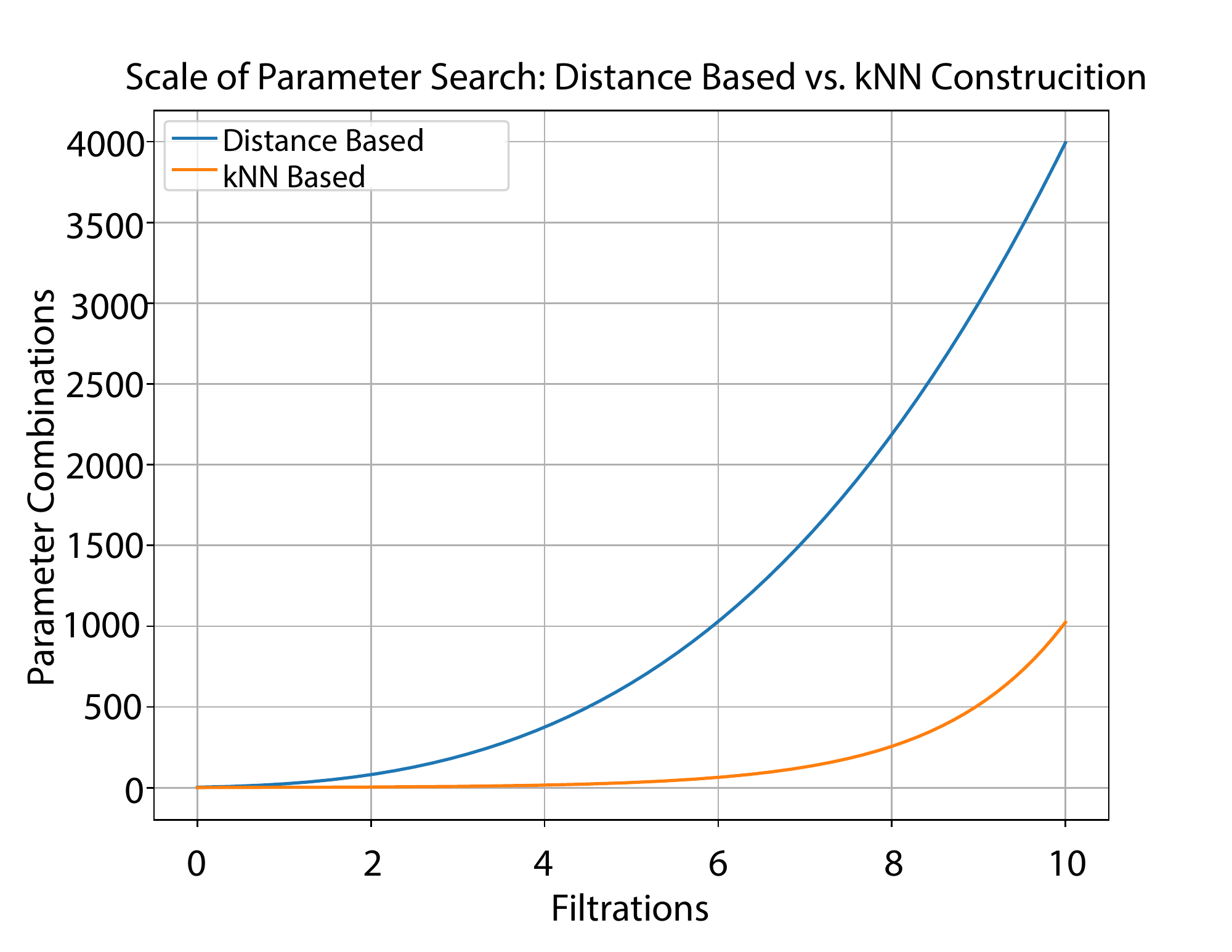}
  \caption{Comparison of the scales of parameter searches necessary between the limited approach to distance based filtrations and the limited approach to kNN-Induced filtrations. }\label{fig:combfig}
\end{figure}

  Ideally, the more standardized filtrations present with kNN Laplacians will reduce the need for parameter optimization entirely, significantly reducing computation and time requirements. As opposed to performing grid search for each dataset, we can universally choose a given set of weights that decrease as connectivity information decreases, such as $\{\zeta_t = 1/ t \}, t = 1,...,p$, and observe whether there is still a meaningful improvement in performance without the need for any parameter search. In case there is still need for some optimization, we can at least significantly restrict the parameter distribution, decreasing the amount of time needed for tuning. Specifically, we weight connectivities as being either unimportant $(\zeta = 0)$, or important $(\zeta = 1)$. This results in the number of tested parameter combinations equaling $(2^{p})(\text{Size of } \gamma \text{ distribution})$. Ultimately, in practice with 8 filtrations we found that this meant fewer than 1/5 the amount of tested parameter combinations were needed to obtain optimal or near optimal results compared to the original construction. In most cases, however, no parameter search was even necessary at all. In Figure \ref{fig:combfig}, we include a chart illustrating the scale of the respective parameter searches as we vary the number of filtrations. 

  We see from this that, for a reasonable number of filtrations, the parameter search necessary for the kNN construction is a fraction of that needed for the standard construction, while the results listed in Table \ref{tab: avg} showcase that the performance is still optimal or at least near optimal compared to other dimensionality reduction techniques. 
 
\begin{figure}[H]
        \centering
        \includegraphics[width=0.75\linewidth]{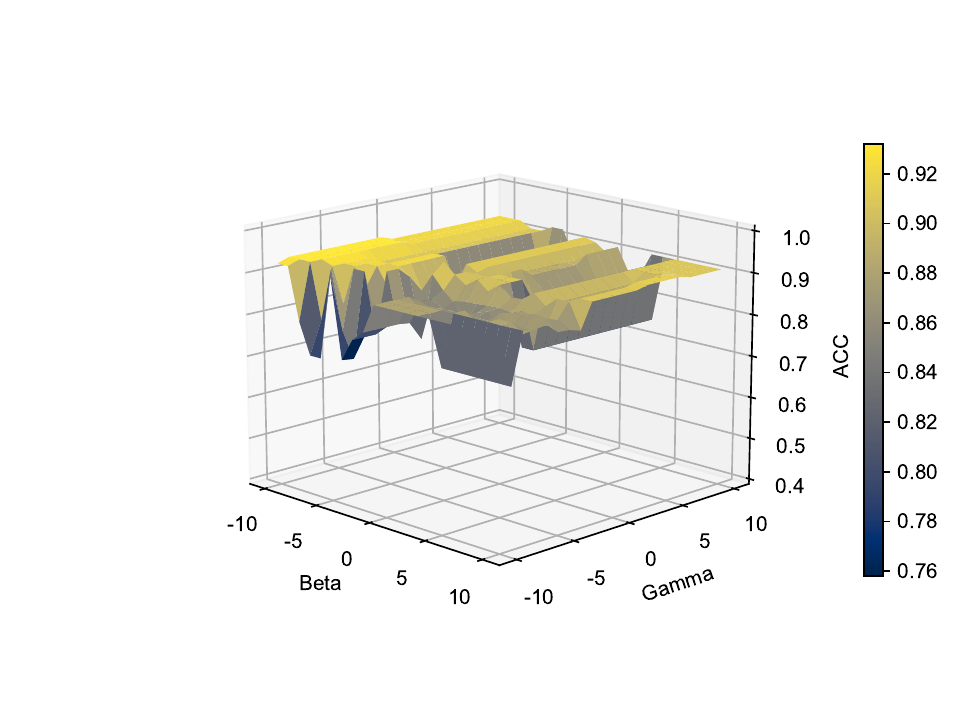}
        \caption{Variations in KMeans accuracy for different combinations of $\gamma$ and $\beta$ parameter values}\label{fig:accfig}
\end{figure}

 The $\beta$ and $\gamma$ parameters are similarly found via grid search. In Figure \ref{fig:accfig}, we depict how different parameter combinations impact the accuracy of our K-Means clustering. Ultimately, parameter values ranging from $10^{-10}$ to $10^{10}$ were found to produce stable results.

\subsection{Visualization of tPCA Eigen-Gene}

In the context of scRNA-seq clustering, an eigen-gene refers to the principal components produced by our dimensionality reduction. Eigen-genes summarize the gene expression patterns within a cluster given that they are linear sums of the features which explain the most variation in that cluster. By reducing our data to two dimensions via UMAP or tSNE, we can visualize our eigen-genes in a 2D plot. This visualization can identify the clusters with similar or distinct gene expression profiles. 

\begin{figure}[H]
        \centering
        \includegraphics[width=\linewidth]{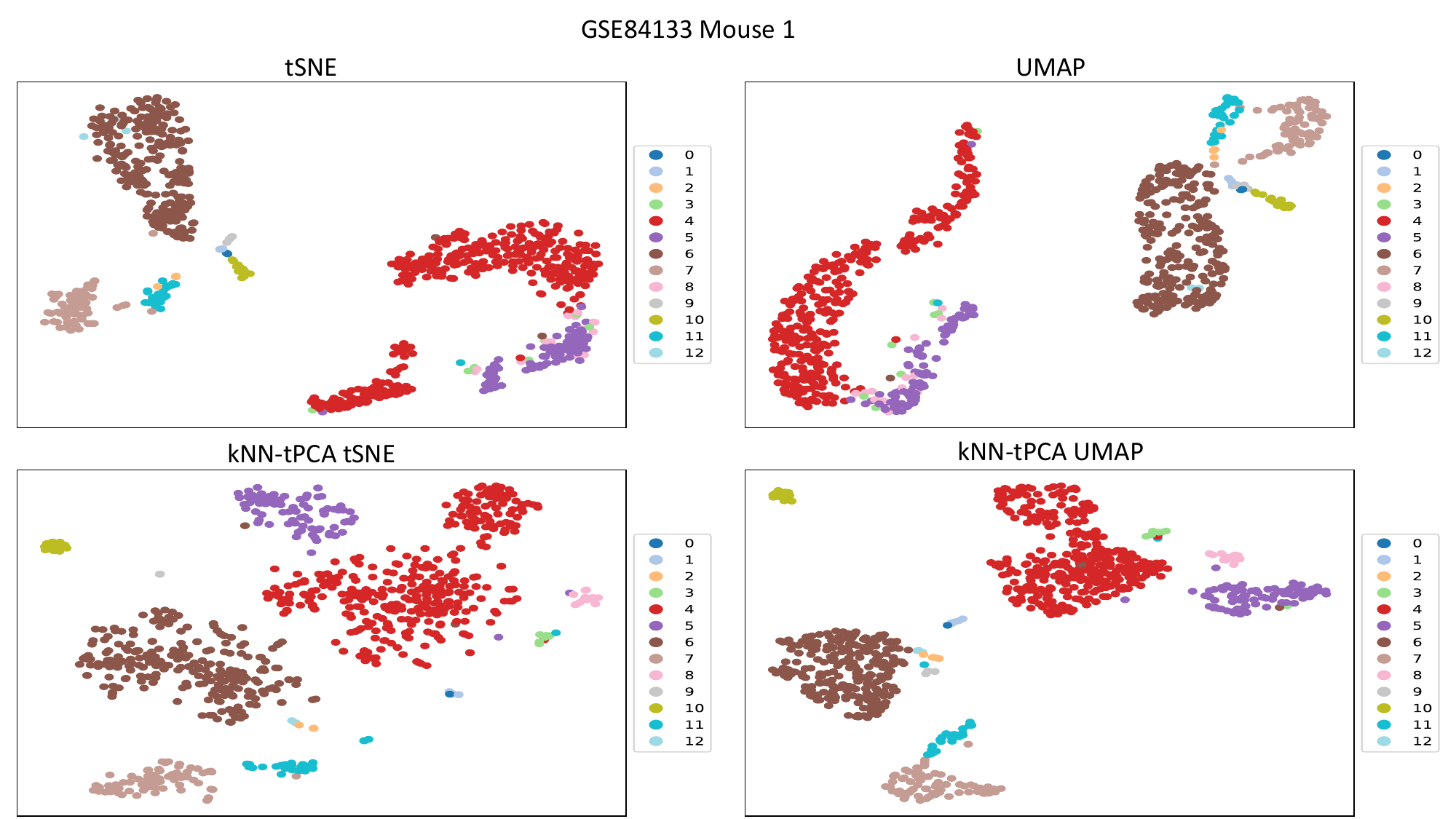}
        \caption{Comparison of visualization techniques between PCA-enhanced tSNE and UMAP, and kNN-tPCA-Enhanced tSNE and UMAP for GSE84133mouse1. Data was log-transformed, with low variance genes removed. For kNN-tPCA-Enhanced tSNE and UMAP, $\zeta$ weights were chosen universally as $\{\zeta_t\} = \{1/t\}$ for the $t$th filtration, and data was reduced to $k=50$ dimensions. Cells are color coded according to true cell types provided by original authors. Labels 0 through 12 correspond to B cells, T cells, Activated Stellate, $\alpha$ cells, $\beta$ cells, $\delta$ cells, Ductal cells, Endothelial cells, $\gamma$ cells, Immune cells (other), Macrophage cells, Quiescent Stellate, and Schwann cells respectively. }\label{fig:eiggenefig1}
\end{figure}

\begin{figure}[H]
        \centering
        \includegraphics[width=\linewidth]{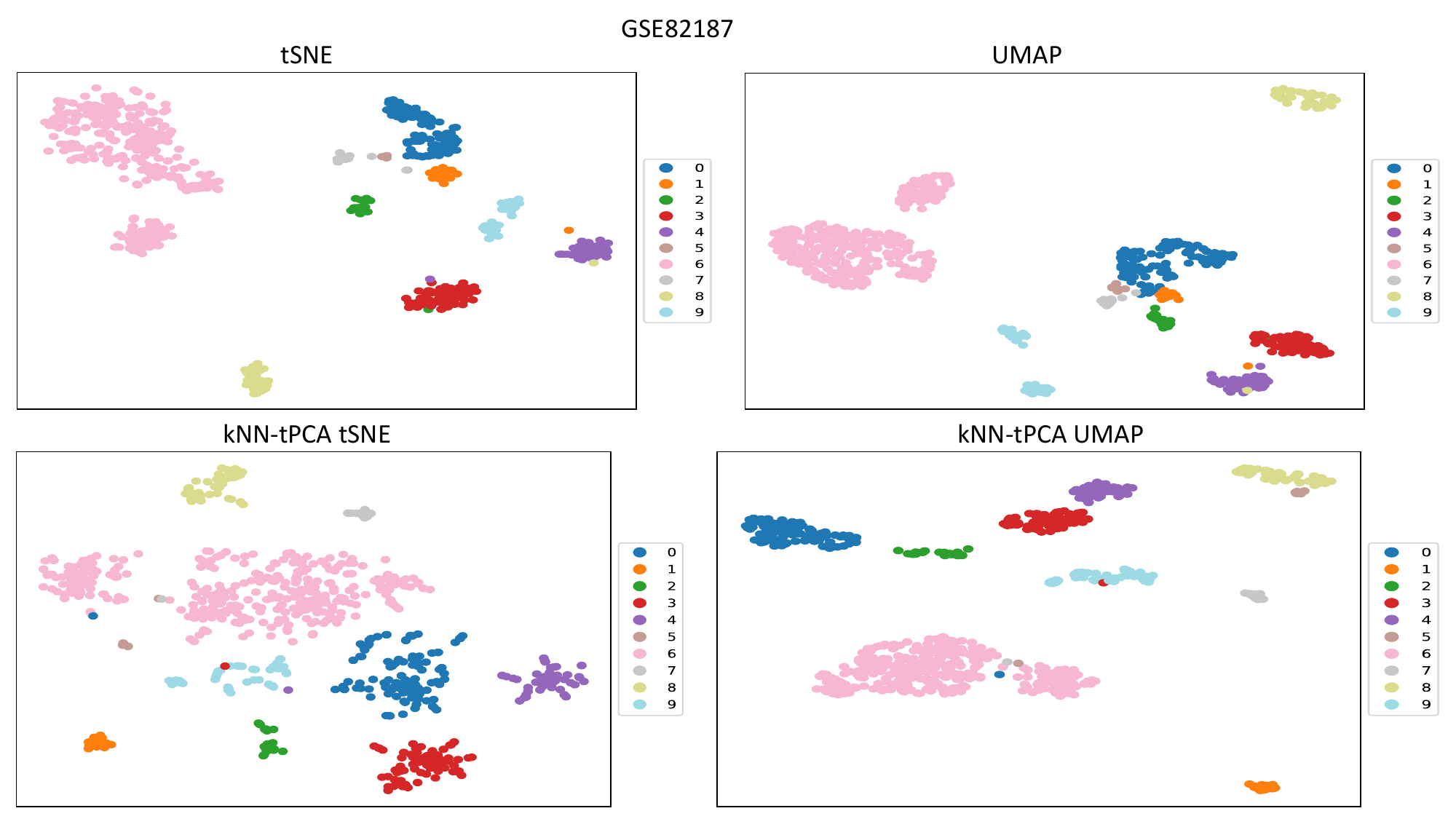}
        \caption{Comparison of visualization techniques between PCA-enhanced tSNE and UMAP, and kNN-tPCA-Enhanced tSNE and UMAP for GSE82187. Data was log-transformed, with low variance genes removed. For kNN-tPCA-Enhanced tSNE and UMAP, $\zeta$ weights were chosen universally as $\{\zeta_t\} = \{1/t\}$ for the $t$th filtration, and data was reduced to $k=50$ dimensions. Cells are color coded according to true cell types provided by original authors. Labels 0 through 9 correspond to Astro cells, Ependy-C cells, Ependy-Sec cells, Macrophage cells, Microglia cells, NSC cells, Neuron cells, OPC cells, Oligo cells, and Vascular cells respectively. }\label{fig:eiggenefig2}
\end{figure}

\begin{figure}[H]
        \centering
        \includegraphics[width=\linewidth]{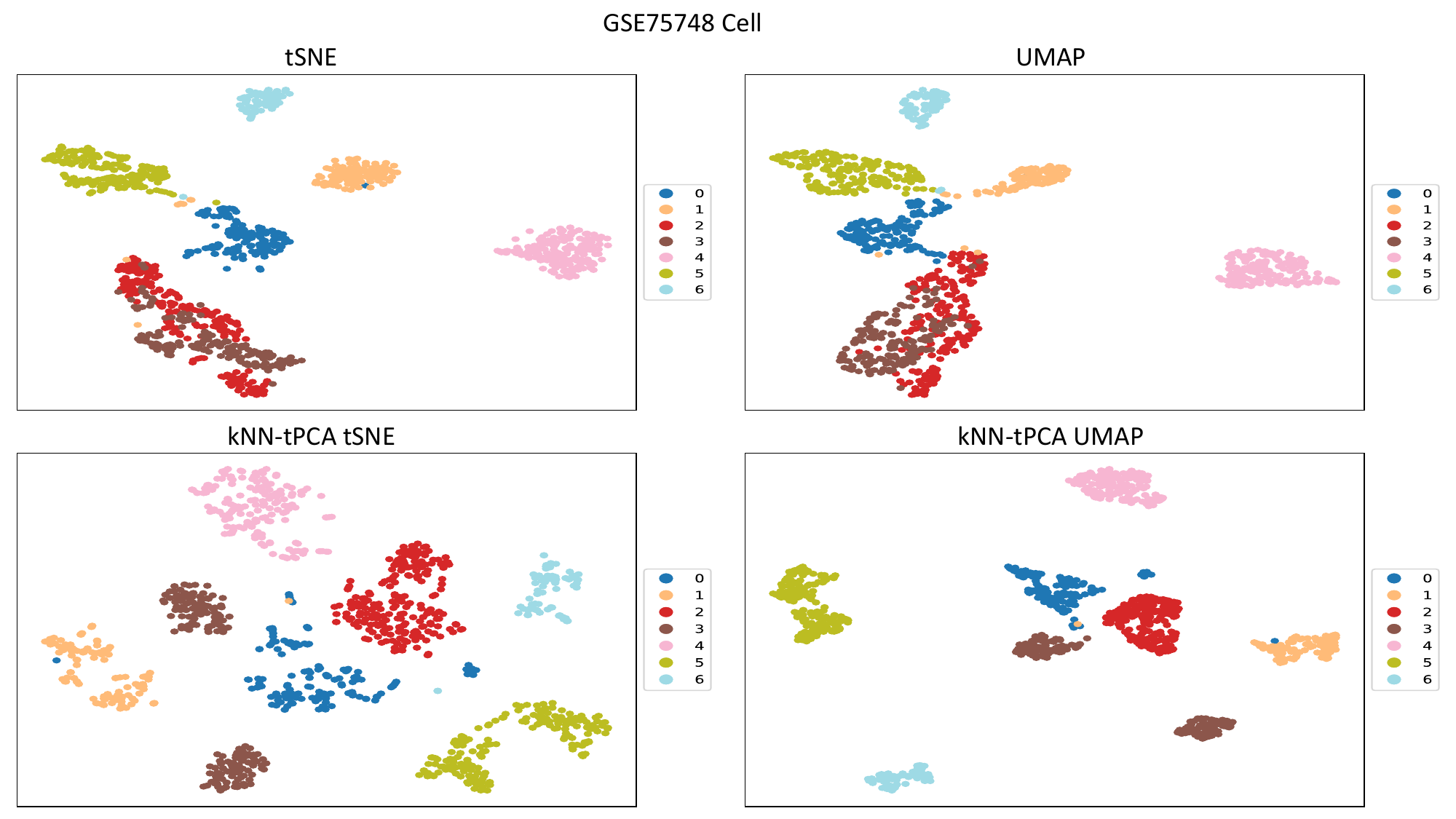}
        \caption{Comparison of visualization techniques between PCA-enhanced tSNE and UMAP, and kNN-tPCA-Enhanced tSNE and UMAP for GSE75748cell. Data was log-transformed, with low variance genes removed. For kNN-tPCA-Enhanced tSNE and UMAP, $\zeta$ weights were chosen universally as $\{\zeta_t\} = \{1/t\}$ for the $t$th filtration, and data was reduced to $k=50$ dimensions. Cells are color coded according to true cell types provided by original authors. Labels 0 through 6 correspond to DEC cells, EC cells, H1 cells, H9 cells, HFF cells, NPC cells, and TB cells respectively.  }\label{fig:eiggenefig3}
\end{figure}

For a dataset containing, say, 20,000 genes, an aggressive reduction to $k=2$ dimensions typically results in poorly maintaining the integrity of the data, leading to ineffective visualizations of the eigen-genes. Thus, an important step in the data visualization process is pre-processing. If we first reduce our data to, say, $k=50$ dimensions via PCA before then reducing again to $k=2$ dimensions via tSNE or UMAP, we should see an improvement in the representation of our clusters. However, given the associated weaknesses with traditional PCA that we have discussed previously, there may be additional benefit to be gained from pre-processing the data with Topological PCA instead. In Figures \ref{fig:eiggenefig1}, \ref{fig:eiggenefig2}, \ref{fig:eiggenefig3} we compare the 2D visualizations for several of the tested datasets when pre-processing via PCA and tPCA to assess this improvement in terms of visualization and potential biological insights.

In Figure \ref{fig:eiggenefig1}, we compare visualization techniques for GSE84133mouse1. In Veres et al, $\beta$ cells were found to have heterogeneity between two distinct subpopulations \cite{article6}. However, traditional PCA-enhanced tSNE separates the subpopulations into two clusters that are far away and considerably mixed 
 with $\delta$ and other cell types. Our method manages to visualize the cells more similarly, while still displaying the genetic heterogeneity in the population. Furthermore, there is considerably improved separation between the $\beta$ cells and other cell types, particularly $\delta$ cells. For UMAP, we observe that PCA-enhanced UMAP clusters all cell types into two relatively homogeneous clusters. Pre-processing with kNN-tPCA, meanwhile, manages to separate all cell types fairly well, with the exception of Endothelial and Quiescent Stellate cells. This is likely explained by quiescent stellate cells being located primarily around vascular cells in the pancreas, including Endothelial cells, leading to similar gene expression profiles between the two cell types. Gaining a further understanding of this spatial organization is crucial for understanding the mechanisms underlying pancreatic diseases. 
In Figure \ref{fig:eiggenefig2}, we compare visualization techniques for GSE82187. For tSNE as well as UMAP, we note improved separation between Astro, Ependy-C, and OPC cells when pre-processing with tPCA rather than traditional PCA. Furthermore, like with traditional PCA-enhanced UMAP and tSNE, kNN-tPCA pre-processing still enables us to identify the distinct D1 and D2 medium spiny neuron subtypes even when inducing sparsenss in our principal components. In both of our improved visualizations, there seems to be more of a continuous gradient between the subtypes rather than a discrete separation. Continuous gradients indicate that neurons within each subtype lie on a spectrum of gene expression values, with many cells having a range of intermediate expression values. These results are supported by the findings in Gokce et al \cite{Gokce2016CellularTO}. 
In Figure \ref{fig:eiggenefig3}, we compare visualization techniques for GSE75748cell. We note for both tSNE and UMAP, the PCA pre-processed version clusters H1 and H9 cells into one homogeneous cluster given the similar gene expression profile of these cells \cite{article5}. However, the kNN-tPCA enhanced versions were still able to differentiate these cell types. The same can be said of DEC and EC cells.  DEC cells were also found to have lower similarity in their clustering with kNN-tPCA enhanced tSNE, indicating a heterogeneous pool of DEC cells. These results are supported by the findings in Chu et al \cite{article5}. In both instances when pre-processing with tPCA, the H9 cells formed two distinct clusters, indicating some kind of possible heterogeneity in the genetic profiles of these cells. 

\subsection{RS Plot Analysis}
\begin{figure}[H]
        \centering
        \includegraphics[width=0.8\linewidth]{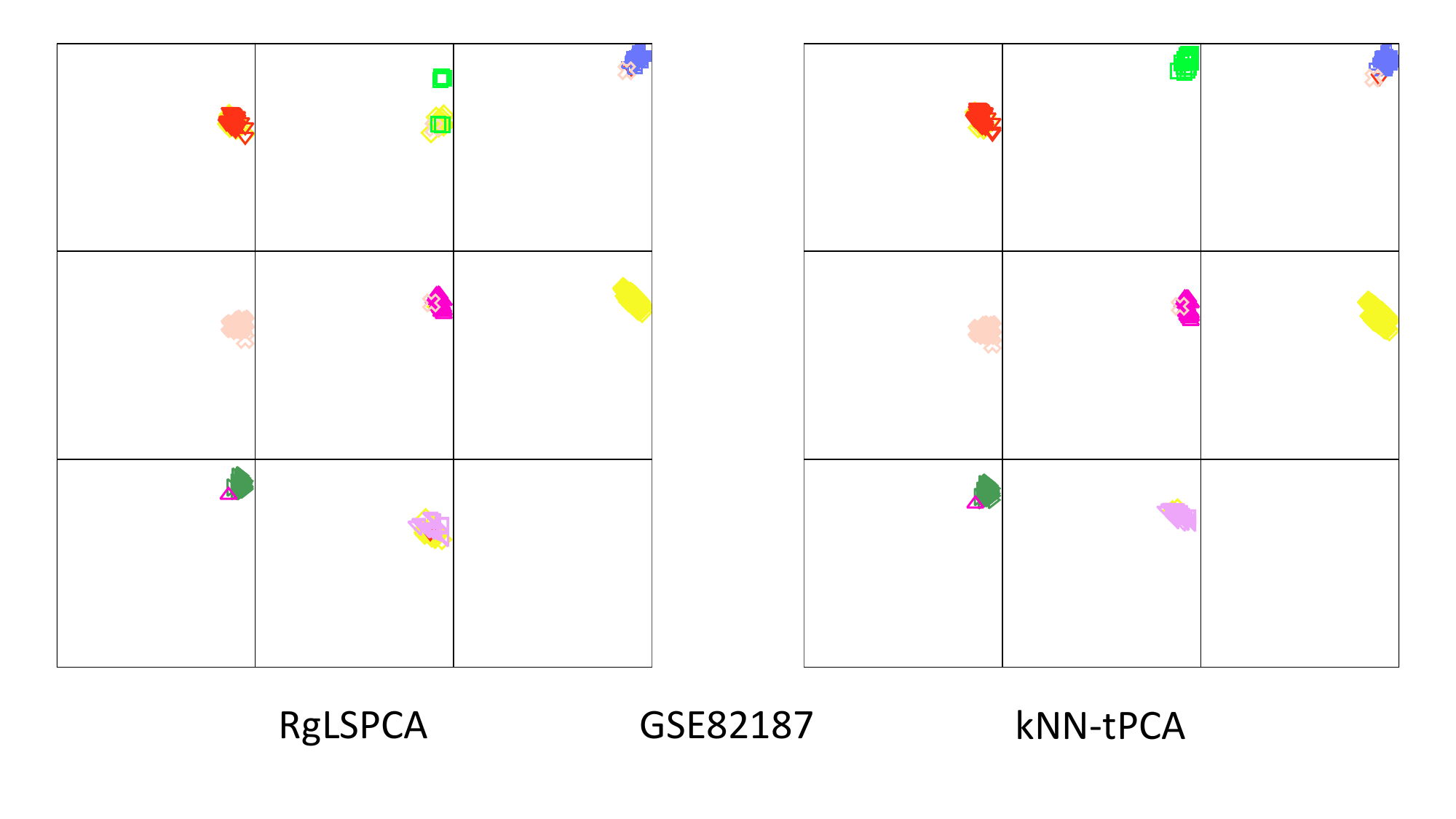}
        \caption{RS plots of clusters generated from RgLSPCA and kNN-tPCA based dimensionality reduction. The $ {x}$-axis is the residual score, and the $ {y}$-axis is the similarity score. Each section corresponds to one cluster and the data were colored according to the predicted labels from kNN on the GSE82187 dataset at $k=100$.}\label{fig:rs}
\end{figure}

To more effectively visualize our gene expression data after dimensionality reduction, we can generate Residue-Similarity plots for some of the tested datasets \cite{hozumi2022ccp}. We can then compare results for classifying cell types after reducing the data via RgLSPCA and kNN-tPCA. In Figure \ref{fig:rs} we produce RS plots for each method on GSE82187 to compare classification accuracy. We observe a significant improvement, particularly in identifying the cell types in panels two and eight, or Ependy-C and Vascular cells respectively. Note specifically that for Ependy-C cells the samples are situated in the top-right corner, indicating a significantly improved cluster boundary separation and inter-cluster similarity in that clustering when utilizing persistent Laplacian regularization. 

Similarly, for GSE67835 we note a considerable improvement in our ability to correctly identify replicating fetal neurons and Microglia in panels five and six respectively. For Microglia cells in particular, we again observe a significant improvement in the residual score for that clustering, indicating that tPCA yields a greater dissimilarity between these cells and other cell types than RgLSPCA. Specifically, tPCA improves the separation between Microglia and quiescent fetal neurons/OPC cells. In panel one, we note that classification after dimensionality reduction via kNN-tPCA has a slightly greater tendency to misidentify OPC cells with lower similarity scores as Microglia cells, indicating that these cells exhibited a similar gene expression profile, which is supported by the findings in Darmanis et. al. for a subset of the OPC population \cite{Darmanis2015ASO}. 
\begin{figure}[H]
        \centering
        \includegraphics[width=0.8\linewidth]{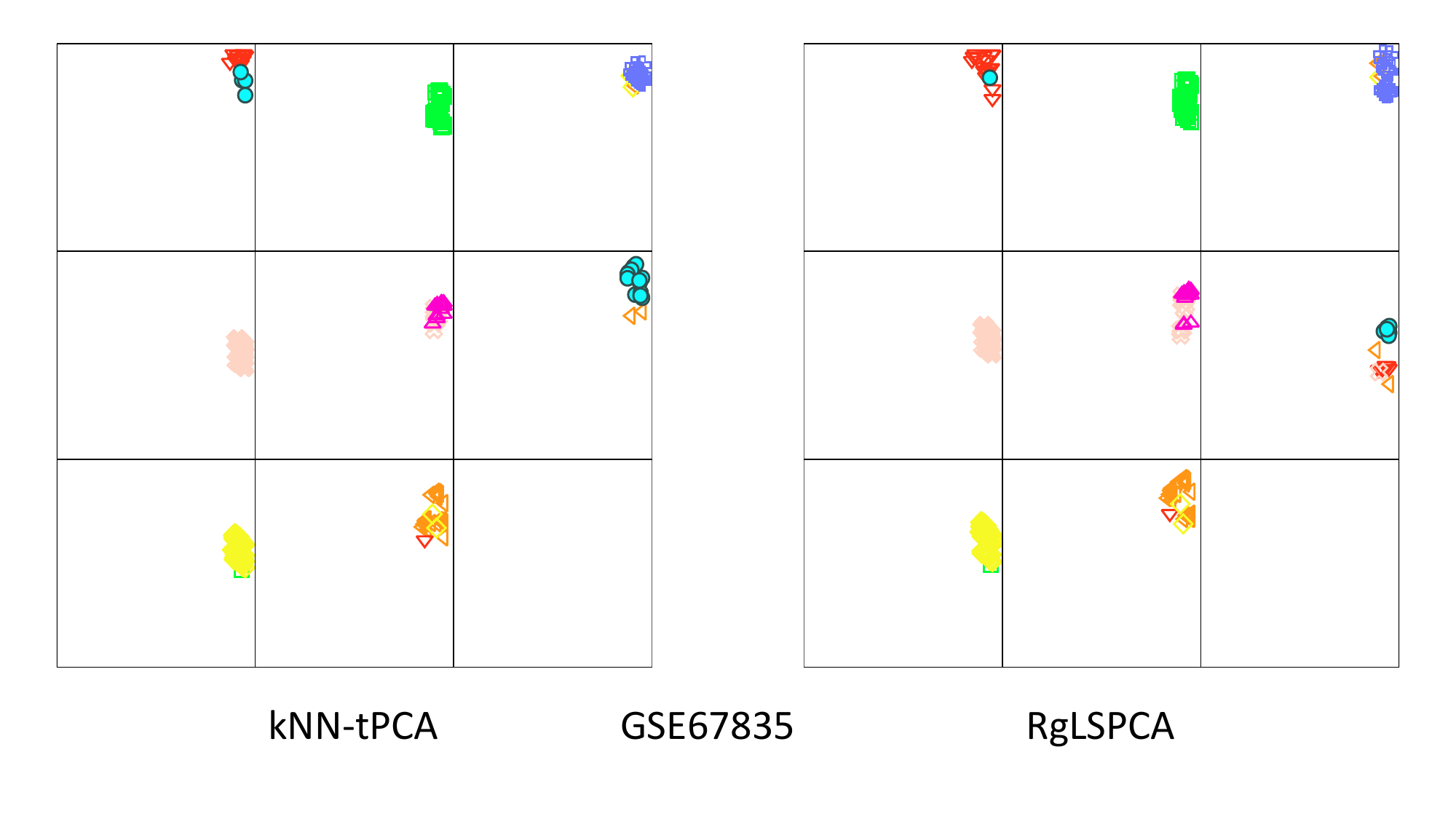}
        \caption{RS plots of clusters generated from RgLSPCA and kNN-tPCA based dimensionality reduction. The $ {x}$-axis is the residual score, and the $ {y}$-axis is the similarity score. Each section corresponds to one cluster and the data were colored according to the predicted labels from KNN on the GSE67835 dataset at $k=100$.}\label{fig:rs2}
\end{figure}

\section{Conclusion}
Single Cell RNA sequencing technologies have grown considerably in popularity in recent years, and with the ability to reveal vast amounts of information regarding the drivers behind various diseases as well as potential bio-therapeutic targets, effective analysis of this data is of paramount importance in the field of biomedical research. As we have seen, intrinsic high dimensionality of the data introduces computational complexity as well as considerable noise, hindering any meaningful analysis. Thus, dimensionality reduction is a crucial step of the process, and we seek, as always, to maximize the accurate representation of our data in the new, reduced space. To this end, we propose topological PCA for scRNA-seq clustering. This method combines a new robustness via L$_{2,1}$ norm regularization, sparsity constraints, and improved geometrical structure capture via persistent Laplacian regularization. 

Extensive benchmark testing on 11 scRNA-seq datasets showcases that our proposed method significantly outperforms other similar PCA enhancements, as well as non-negative matrix factorization, for KMeans clustering after dimensionality reduction. While previous methods such as graph Laplacian Sparse PCA account for sparsity and local geometry preservation, the method is limited by analysis of a simplicial complex at only a single scale. Furthermore, Frobenius norm regularization is sensitive to outliers. The incorporation of a persistent Laplacian term contributes to multi-scale analysis through a sequence of filtrations, as well as persistent homology information derived from the harmonic spectra of our Laplacian matrices. Compared to NMF, we observe an average improvement of 13.53\% for NMI and 31.92\% for ARI. 

While our method achieves superb results for clustering analysis after dimensionality reduction, there is still considerable room for improvement. First, our method considers only $\mathcal{L}_0$ Laplacian, and therefore lacks higher order connectivity information. Furthermore, there remains the work of continuing our parameter analysis, to arrive at a means of optimizing our $\{\zeta\}$ weights which is more efficient and produces more optimal results than simple grid search. While kNN-induced persistent Laplacians seem to be less dependent on parameter tuning, there is still added benefit to examining means of optimizing the performance, and so we should hope to achieve this in a more efficient manner. 

\section{Data and Model Availability}
The data and model used to produce these results can be obtained at the Single Cell Data Processing and RpLSPCA scRNA-seq GitHub Repositories: 

\href{https://github.com/seanfcottrell/Topological-PCA}{Topological PCA} GitHub repository: 
https://github.com/seanfcottrell/Topological-PCA 

\href{https://https://github.com/hozumiyu/SingleCellDataProcess}{Single Cell Data Processing} GitHub repository: 
https://github.com/hozumiyu/SingleCellDataProcess

\section{Acknowledgments}
This work was supported in part by NIH grants R01GM126189, R01AI164266, and R35GM148196, National Science Foundation grants DMS2052983, DMS-1761320, and IIS-1900473, NASA  grant 80NSSC21M0023,   Michigan State University Research Foundation, and  Bristol-Myers Squibb  65109.

\clearpage 

\bibliographystyle{unsrt}
\bibliography{refs}

\end{document}